\documentclass[referee]{raa}

\usepackage{graphicx,times}
\usepackage{natbib}
\usepackage{amssymb,amsmath}
\bibpunct{(}{)}{;}{a}{}{,}

\usepackage[pagebackref=true]{hyperref}

\begin{document}

   \title{Constraining Lyman–Werner Feedback from Velocity Acoustic Oscillations in the Cosmic Dawn 21 cm Signal}

 \volnopage{ {\bf 20XX} Vol.\ {\bf X} No. {\bf XX}, 000--000}
   \setcounter{page}{1}

   \author{Xi Du
      \inst{1,2}
      \href{https://orcid.org/0009-0004-5379-2248}{\includegraphics[scale=0.05]{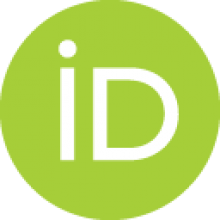}}
   \and Bin Yue
      \inst{1}$^*$
      \href{https://orcid.org/0000-0002-7829-1181}{\includegraphics[scale=0.05]{orcid-ID.png}}
   \and Furen Deng
      \inst{1,2}
      \href{https://orcid.org/0000-0001-8075-0909}{\includegraphics[scale=0.05]{orcid-ID.png}}
    \and Yidong Xu
      \inst{1}
      \href{https://orcid.org/0000-0003-3224-4125}{\includegraphics[scale=0.05]{orcid-ID.png}}
    \and Yan Gong
      \inst{1,2}
      \href{https://orcid.org/0000-0003-0709-0101}{\includegraphics[scale=0.05]{orcid-ID.png}}
    \and Ely D. Kovetz
      \inst{3}
      \href{https://orcid.org/0000-0001-9256-1144}{\includegraphics[scale=0.05]{orcid-ID.png}}
    \and Xuelei Chen
      \inst{1,2}
      \href{https://orcid.org/0000-0001-6475-8863}{\includegraphics[scale=0.05]{orcid-ID.png}}
       \footnotetext{$*$ Corresponding author}
   }

   \institute{ State Key Laboratory of Radio Astronomy and Technology, National Astronomical Observatories, Chinese Academy of Sciences,
             Beijing 100101, China; 
             {\it yuebin@nao.cas.cn}\\
        \and
             School of Astronomy and Space Science, University of Chinese Academy of Sciences,
             Beijing 100049, China\\
        \and
             Physics Department, Ben-Gurion University of the Negev,
             Beersheba, Israel\\
\vs \no
   {\small Received 20XX Month Day; accepted 20XX Month Day}
}

\abstract{During Cosmic Dawn, Pop III stars could be formed in minihalos through molecular hydrogen (H$_2$) cooling. The minimum halo mass required for H$_2$ cooling is highly sensitive to Lyman–Werner (LW) radiation, which dissociates H$_2$ and regulates star formation. However, the efficiency of LW feedback remains poorly constrained due to the lack of direct observations of Pop III stars.
The dark matter–baryon relative streaming velocity suppresses star formation in low-mass halos and imprints characteristic Velocity Acoustic Oscillation (VAO) features in the 21 cm power spectrum. These features are particularly sensitive to the cooling threshold mass: if LW feedback raises the minimum halo mass above the streaming-sensitive regime, the VAO signal is strongly suppressed. This makes the VAO wiggles a promising indirect probe of LW feedback during  Cosmic Dawn.
We investigate the feasibility of constraining LW feedback parameters using semi-numerical 21 cm lightcone simulations.  We compute the multi-frequency angular power spectrum (MAPS) to isolate the VAO features and train a Convolutional Neural Network (CNN) to infer the LW feedback efficiency and the baseline cooling threshold.
 We find that in the absence of instrumental noise, the LW feedback efficiency can be accurately recovered from the VAO features. 
 However, for the SKA-low AA* configuration, meaningful constraints require integration times exceeding $10^4$ hours under optimistic foreground assumptions.
  Nonetheless, our results demonstrate that VAO features provide a physically robust and potentially powerful probe of LW feedback at  Cosmic Dawn.
\keywords{(cosmology:) dark ages, reionization, first stars -- (cosmology:) large-scale structure of Universe -- (galaxies:) intergalactic medium -- galaxies: high-redshift 
}
}

   \authorrunning{Du et al. }            
   \titlerunning{LW feedback \& VAO}  
   \maketitle

%
\section{Introduction}           
\label{sect:intro}

The Cosmic Dawn epoch ($z\sim 30-10$) marks the emergence of the first generation of stars, commonly referred to as Population III (Pop III) stars. These stars are believed to form in minihalos with virial temperatures $T_{\rm vir} \gtrsim 10^3$ K, where cooling is enabled by molecular hydrogen (H$_2$) (e.g. \citealt{barkanaBeginningFirstSources2001,gloverOpenQuestionsStudy2008,brommFormationFirstStars2009,gloverFirstStars2013,annurev:/content/journals/10.1146/annurev-astro-071221-053453}). The formation efficiency and abundance of Pop III stars therefore depend critically on the minimum halo mass required for H$_2$ cooling.

A key regulatory mechanism during this era is Lyman–Werner (LW) radiation. Photons in the LW band (11.2 - 13.6 eV) dissociate H$_2$ via the two-step Solomon process, suppressing cooling and raising the critical halo mass for star formation \citep{Field1966ARA&A...4..207F,
stecherPhotodestructionHydrogenMolecules1967}.  
As Pop III stars themselves are sources of LW radiation, this introduces a self-regulating feedback loop: early star formation generates LW radiation, which subsequently suppresses further star formation in low-mass halos (\citealt{kulkarniCriticalDarkMatter2021,safranek-shraderStarFormationFirst2012,xuLATEPOPIII2016,proleDarkMatterHalos2023}). The strength of this feedback determines the star formation history of the smallest halos and influences the thermal and radiative evolution of the intergalactic medium (IGM) (\citealt{gloverFirstStars2013,cruzEffectiveModel21cm2025}). Despite its importance, the efficiency of LW feedback remains poorly constrained due to the absence of direct observations of Pop III stars.

The redshifted 21 cm signal of neutral hydrogen provides a unique observational window into  Cosmic Dawn  (\citealt{furlanettoCosmologyLowFrequencies2006,Chen2008,gessey-jonesDeterminationMassDistribution2025,bevinsAstrophysicalConstraintsSARAS2022,dhandhaExploitingSynergiesJWST2025}). The 21 cm brightness temperature traces the thermal and radiative state of the IGM, which is shaped by the first stars through Ly$\alpha$ coupling, X-ray heating, and LW feedback. In particular, the formation of Pop III stars in minihalos leaves imprints on the large-scale structure of the 21 cm signal (\citealt{Mirocha2018MNRAS.478.5591M,venturaRolePopIII2022,venturaSemianalyticalModellingPop2025,tanakaModellingPopulationIII2021,gessey-jonesDeterminationMassDistribution2025}).

An additional ingredient relevant to early star formation is the dark matter–baryon relative streaming velocity, generated at recombination \citep{tseliakhovichSuppressionSpatialVariation2011}. This coherent supersonic velocity suppresses gas accretion in low-mass halos and modulates star formation on large scales (\citealt{fialkovImpactRelativeMotion2012,greifDelayPopulationIII2011,2026ApJ...997..202W,tanakaEffectBaryonicStreaming2013,schauerDwarfGalaxyFormation2023}). The resulting spatial modulation produces characteristic Velocity Acoustic Oscillation (VAO) features in the 21 cm power spectrum \citep{dalalLargescaleBAOSignatures2010,munozRobustVelocityinducedAcoustic2019,mcquinnIMPACTSUPERSONICBARYONDARK2012,tseliakhovichRelativeVelocityDark2010,ali-haimoudNewLight212014}. 
These features appear at well-defined comoving scales inherited from pre-recombination acoustic physics and are strongest when star formation occurs in halos whose circular velocities are comparable to the streaming velocity.
Crucially, VAO features are  highly sensitive to the cooling threshold mass. If LW feedback significantly raises the minimum halo mass for star formation, the affected halos become less sensitive to streaming suppression, and the VAO signal is correspondingly weakened or erased. Therefore, the amplitude and redshift evolution of VAO wiggles encode information about the efficiency of LW feedback.

Previous works have explored VAO signatures in the 21 cm power spectrum and their prospects as a probe of Pop III stars \citep{Zhang_2024}, dark matter properties \citep{hotinliProbingUltralightAxions2022}, and cosmology (\citealt{munozStandardRulerCosmic2019,sarkarMeasuringCosmicExpansion2023,hotinliProbingCompensatedIsocurvature2021}). 
 However, the potential of VAO features to constrain astrophysical feedback parameters—particularly LW feedback—has not been systematically investigated. Moreover, during the Cosmic Dawn, the 21 cm signal evolves rapidly along the line of sight, and lightcone effects can significantly modify the observable power spectrum. These effects motivate the use of estimators that preserve frequency evolution, such as the multi-frequency angular power spectrum (MAPS), rather than relying solely on spherically averaged three-dimensional power spectra.

In this work, we investigate whether VAO features in realistic 21 cm lightcone simulations can be used to infer the LW feedback efficiency. We implement a parameterized model of the H$_2$ cooling threshold that incorporates both LW feedback and dark matter–baryon streaming effects, and generate mock lightcone realizations using modified semi-numerical simulations. We extract VAO wiggles from the MAPS and employ a Convolutional Neural Network (CNN) to infer the LW feedback efficiency and the baseline cooling threshold mass. This approach allows us to quantify degeneracies, cosmic variance effects, and observational noise limitations in a controlled framework. 

We find that VAO features provide a robust probe of LW feedback in the absence of instrumental noise. When observational noise corresponding to the SKA-low AA* configuration is included, constraints become significantly weaker, requiring integration times exceeding $10^4$ hours under optimistic assumptions. Nevertheless, our results demonstrate that VAO wiggles offer a physically well-motivated and potentially powerful avenue for probing the cooling physics of the first stars.

  Throughout this paper, we assume a flat $\Lambda$CDM cosmology with   parameters $(\Omega_\Lambda,\Omega_m,\Omega_b,n_s,\sigma_8,h) = (0.692,0.308,0.0484,0.968,0.815,0.678)$ \citep{planckcollaborationPlanck2015Results2016}.

\section{Methods}
\label{sec:method}

\subsection{The Semi-numerical Simulations and the VAO Features}

Similar to \citet{Zhang_2024}, we modified the {\tt 21cmFAST} code (\citealt{mesinger2011MNRAS.411..955M,murray21cmFASTV3Pythonintegrated2020,2025A&A...701A.236D}) to generate the density field $\delta(\vec{r})$ and the dark matter-baryon relative streaming velocity field $\vec{v}_{\rm db}(\vec{r})$, and simulate the evolution of Pop III stars and the LW radiation  field self-consistently. We set a box length of $L_{\rm box} = 1800$ Mpc with $300^3$ cells. Each voxel has side length 6 Mpc. This is larger than the coherence scale of the dark matter-baryon relative streaming velocities which is $\sim3$ Mpc \citep{tseliakhovichRelativeVelocityDark2010}. However, according to \cite{Zhang_2024} (see their Fig. 19), this would just slightly underestimate the amplitude of the VAO wiggles and will not change our conclusion in this paper.
In Appendix \ref{sec:res_boxsize}, we compare the results for different box size and resolution.

\subsubsection{The critical halo mass for H$_2$ cooling and Pop III stars formation}

Pop III stars form in minihalos with $T_{\rm vir} \gtrsim 10^3$ K, because the H$_2$ cooling is efficient only when the gas temperature is $\gtrsim 10^3$ K (\citealt{kashlinskyFormationPopulationIII1983,gloverFirstStars2013}). However, H$_2$ is fragile in front of LW radiation from other Pop III stars, therefore, the critical mass could be larger as only massive minihalos can protect their central H$_2$ from being dissociated by external LW flux \citep{machacekSimulationsPregalacticStructure2001,Wise2007ApJ,OShea2008ApJ,visbalHighredshiftStarFormation2014,kulkarniCriticalDarkMatter2021,schauerInfluenceStreamingVelocities2021}.

To quantify the H$_2$ cooling threshold and the LW feedback efficiency, following the form first proposed by \citet{machacekSimulationsPregalacticStructure2001} and confirmed by \citet{Wise2007ApJ,OShea2008ApJ,fialkov21cmSignatureFirst2013} in their simulations, in this paper we employ a two-parameter formula  to describe the critical minihalo mass allowing H$_2$ cooling and Pop III stars formation in the presence of the LW radiation, 
\begin{equation}
M_{\rm cool1}=M_{\rm cool0}\left[1+\alpha_{\mathrm LW}(4\pi J_{\mathrm LW})^{0.47}\right],
\label{eq:Mcool1}
\end{equation}  
where $J_{\rm LW}$ is the specific intensity of LW radiation, in units of $\mathrm{10^{-21}\ erg\ s^{-1}\ cm^{-2}\ Hz^{-1}\ sr^{-1}}$, and $M_{\rm cool0}$ and $\alpha_{\rm LW}$ are two free parameters.
$M_{\rm cool0}$ is the baseline cooling threshold in the absence of both LW radiation and dark matter-baryon relative streaming velocity.
Typically it corresponds to a virial temperature $T_{\rm vir}\sim 1000$ K ($\sim 10^5~M_\odot$ in Cosmic Dawn), above which H$_2$ cooling becomes efficient (e.g. \citealt{tegmarkHowSmallWere1997,yoshidaSimulationsEarlyStructure2003,abelFormationFragmentationPrimordial2000,barkanaBeginningFirstSources2001}). 
In principle $M_{\rm cool0}$ may depend on redshift \citep{visbalHighredshiftStarFormation2014}, however, to avoid complicating the parameterization, we treat it as a constant parameter. $\alpha_{\rm LW}$ describes the sensitivity of the critical mass to LW radiation, and \citet{fialkov21cmSignatureFirst2013} derived a value 6.96 from simulations. The index number 0.47 in Eq. (1) is from \citet{machacekSimulationsPregalacticStructure2001} and we keep it.

Suppose the circular velocity corresponding to $M_{\rm cool1}$ is  \citep{barkanaBeginningFirstSources2001}, 
\begin{equation}
V_{\rm cool1}=\left(\frac{GM_{\rm cool1}}{R_{\rm vir}(M_{\rm cool1})}\right)^{1/2},
\end{equation}
where $G$ is the gravitational constant and $R_{\rm vir}$ is the virial radius of $M_{\rm cool1}$, then the new critical circular velocity in the presence dark matter-baryon relative streaming velocity $v_{\rm db}$ is \citep{fialkovImpactRelativeMotion2012}
\begin{equation}
V_{\rm cool2}=[V^2_{\rm cool1}+(\alpha_{v_{\rm db}}  v_{\rm db})^2]^{1/2},
\end{equation}
where $\alpha_{v_{\rm db}}=4.015$.
The virial mass of the above circular velocity, $M_{\rm cool2}$, solved from 
\begin{equation}
M_{\rm cool2}=\frac{ R_{\rm vir}(M_{\rm cool2}) V^2_{\rm cool2}}{G},
\label{eq:Mcool2}
\end{equation}
is the critical mass  of minihalos allowing H$_2$ cooling and Pop III stars formation in the presence of both LW radiation and the dark matter-baryon relative streaming motion. 
In our work, LW radiation depends on both position $\vec{r}$ and redshift $z$, therefore $M_{\rm cool1}$ and $M_{\rm cool2}$ are all functions of them. $J_{\rm LW}(\vec{r},z)$, $M_{\rm cool2}(\vec{r},z)$ and the formation of Pop III stars are computed self-consistently in the semi-numerical simulations.

To focus on investigating the feasibility of inferring the LW feedback efficiency, throughout this paper, we fix other Pop III stars characteristic parameters, the star formation efficiency $f_*=0.01$ and the number of X-ray photons per stellar mass $\zeta_X=5\times 10^{55}~M_{\odot}^{-1}$. We limit our investigation in Cosmic Dawn, where the cosmic radiation field is dominated by Pop III stars, so the evolution of $M_{\rm cool2}$ and LW radiation field must be solved self-consistently. Moreover, it is generally believed that most of the IGM is still neutral in Cosmic Dawn, we therefore ignore the reionization process.

\subsubsection{An intuitive estimation of the influence of $M_{\rm cool0}$ and $\alpha_{\rm LW}$ on Pop III stars formation}

From Eq. (\ref{eq:Mcool1}), we see that if $J_{\rm LW}$ is constant, then $M_{\rm cool0}$ would be fully degenerate with $\alpha_{\rm LW}$. However, as we have pointed out, $J_{\rm LW}$ depends on both $\vec{r}$ and $z$, therefore $M_{\rm cool0}$ and $\alpha_{\rm LW}$ can have distinguishable behaviors. 
Before simulating the Pop III stars formation and LW radiation field $J_{\rm LW}(\vec{r},z)$ self-consistently, we first assume an uniform LW background that evolves with redshift, and intuitively estimate  how the parameters $M_{\rm cool0}$ and $\alpha_{\rm LW}$ shape the Pop III stars formation, under the modulation of the dark matter-baryon relative streaming motion. Both $M_{\rm cool0}$ and $\alpha_{\rm LW}$ can boost the $M_{\rm cool1}$, however, the role played by $\alpha_{\rm LW}$ relies on LW radiation, therefore may break the degeneracy. 

Assuming an LW background $J_{\rm LW}^{\rm V14}(z)=J_010^{-\frac{(z-z_0)}{\sigma_{\rm LW}}}$ \citep{visbalHighredshiftStarFormation2014}, we demonstrate three illustrations: (1) $M_{\rm cool0}=2.4\times 10^6~M_\odot$ and $\alpha_{\rm LW}=0$. In this illustration $M_{\rm cool1}$ is independent of redshift and LW radiation. (2) $M_{\rm cool0}=10^5~M_\odot$, $\alpha_{\rm LW}=7$ and $\sigma_{\rm LW}=10$. In this illustration  $M_{\rm cool1}$ is quite sensitive to LW radiation however LW radiation evolves slowly. (3) Similar to (2) but $\sigma_{\rm LW}=1$. In this illustration LW radiation and $M_{\rm cool1}$ evolves rapidly. We adopt these parameter values so that in all illustrations  we have  $M_{\rm cool1}=2.4\times 10^6~M_\odot$ at $z=20$, therefore the results can be compared directly.
 
\begin{figure*}
    \centering
\includegraphics[width=0.25\linewidth]{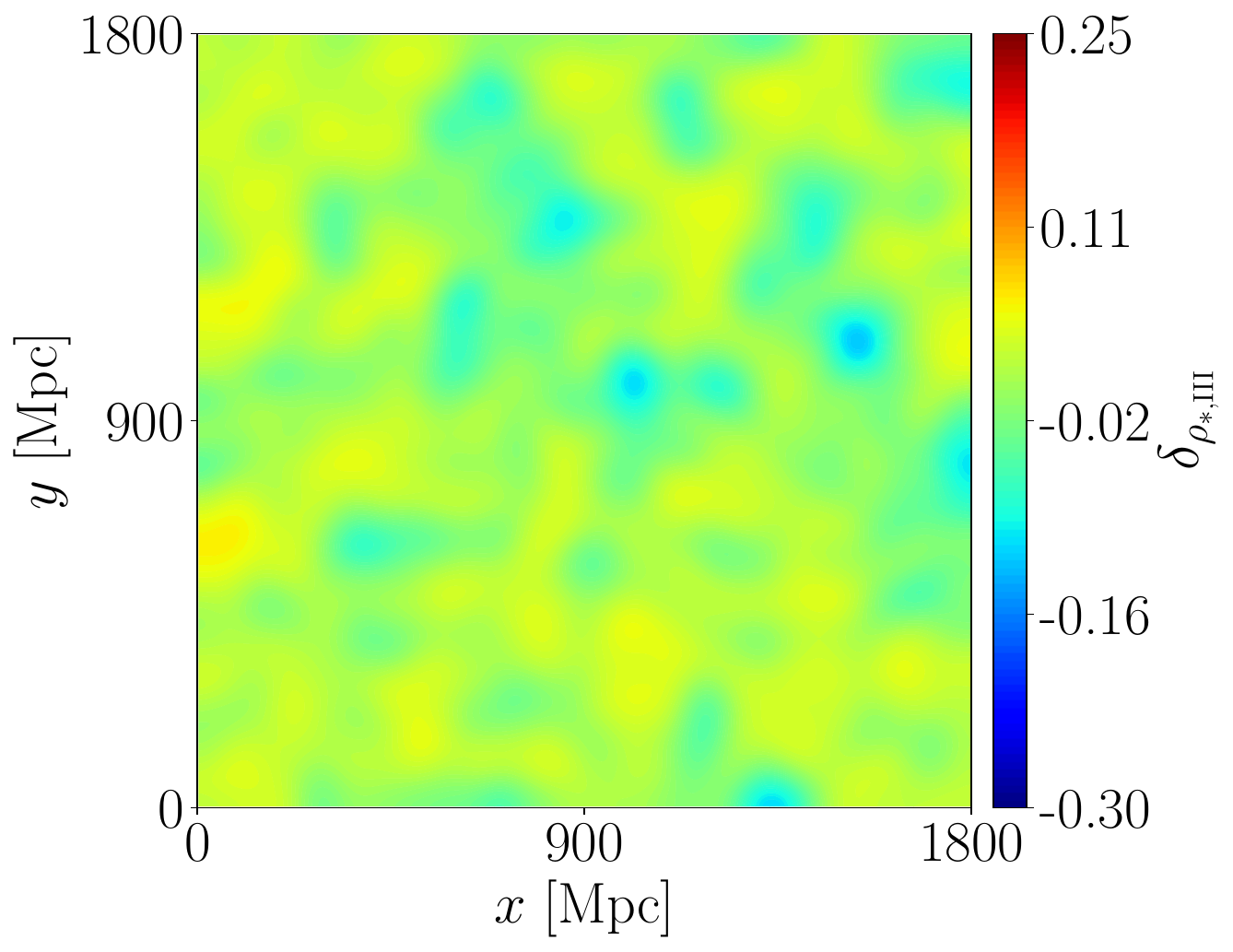}   
    \includegraphics[width=0.25\linewidth]{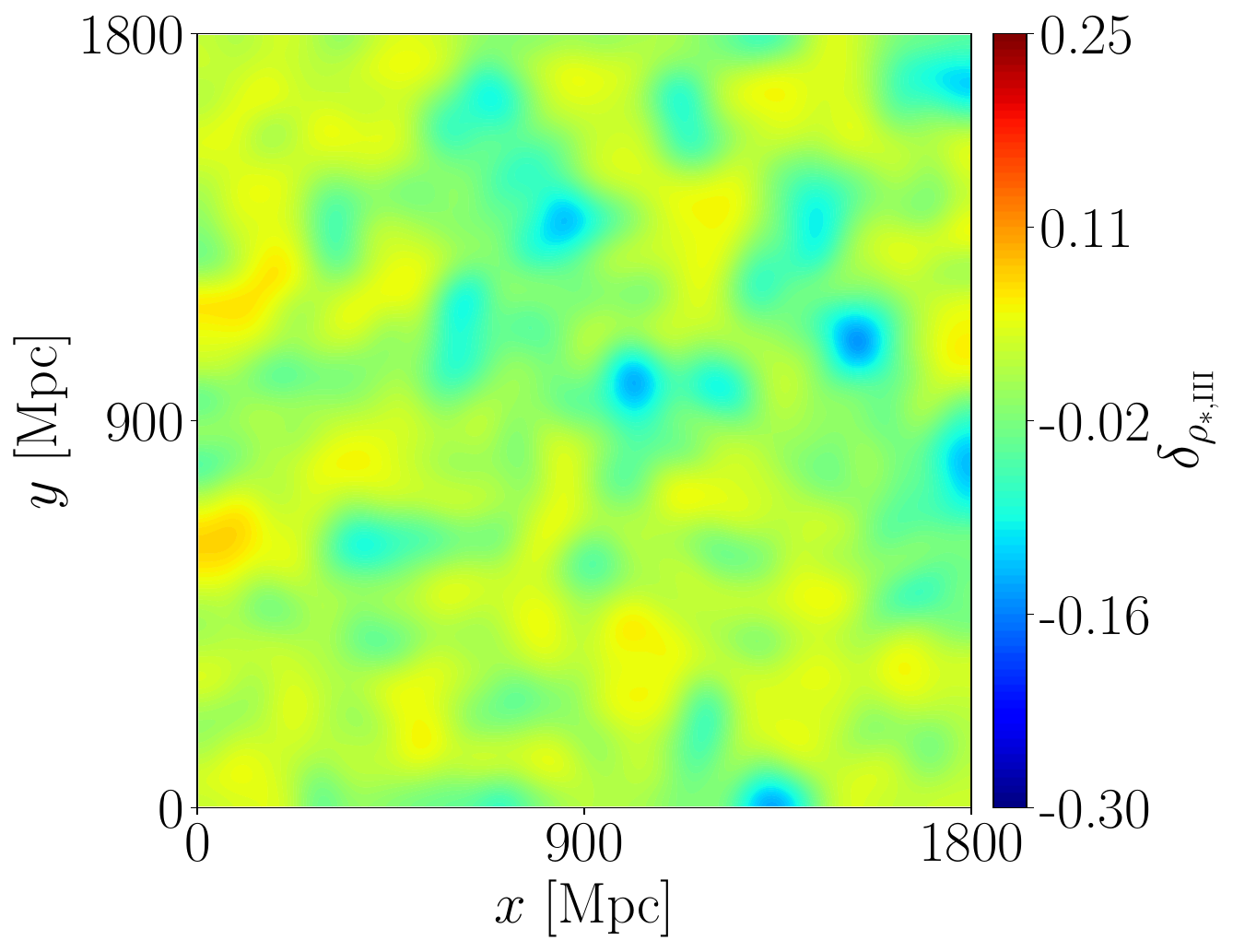}  
\includegraphics[width=0.25\linewidth]{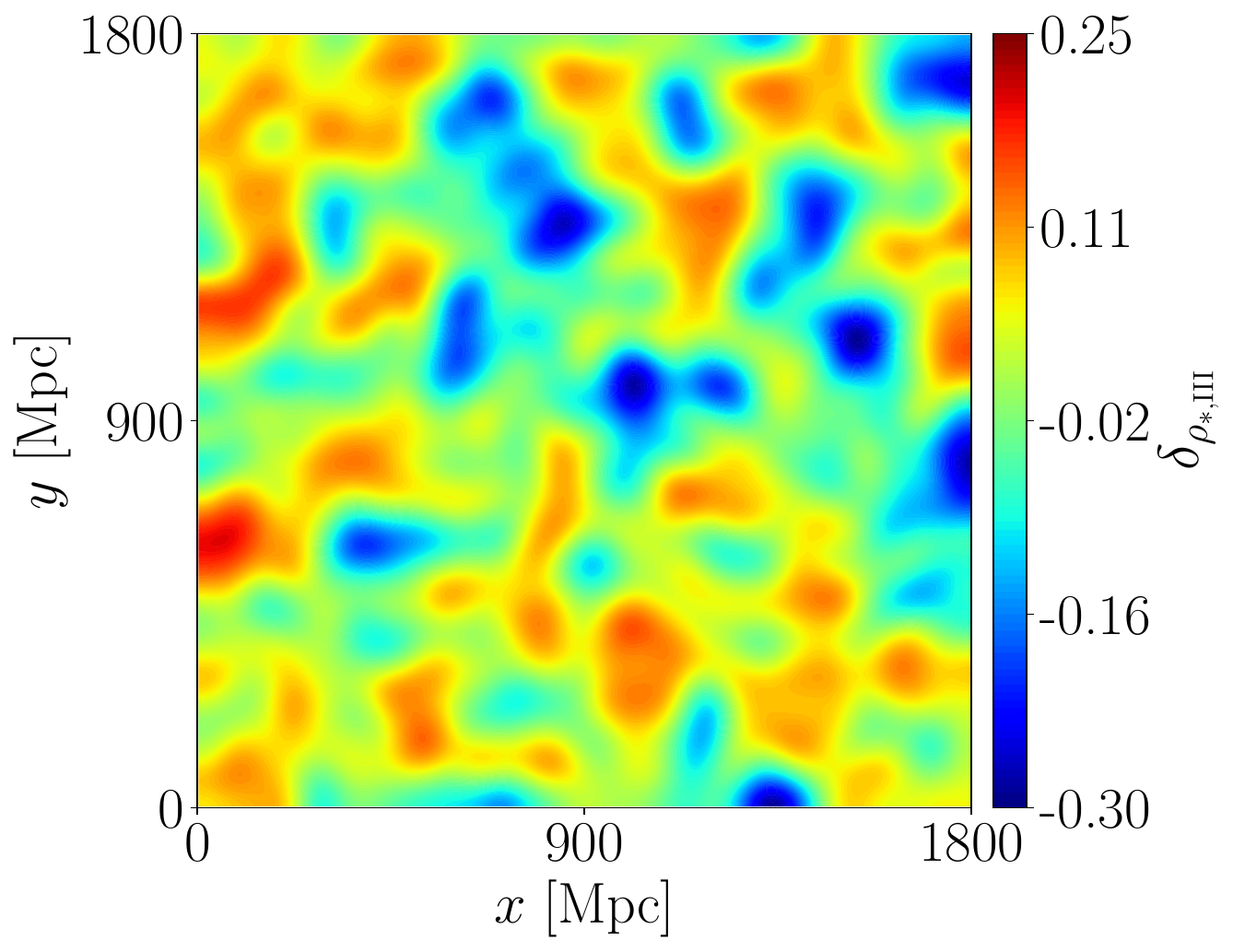}  
\includegraphics[width=0.25\linewidth]{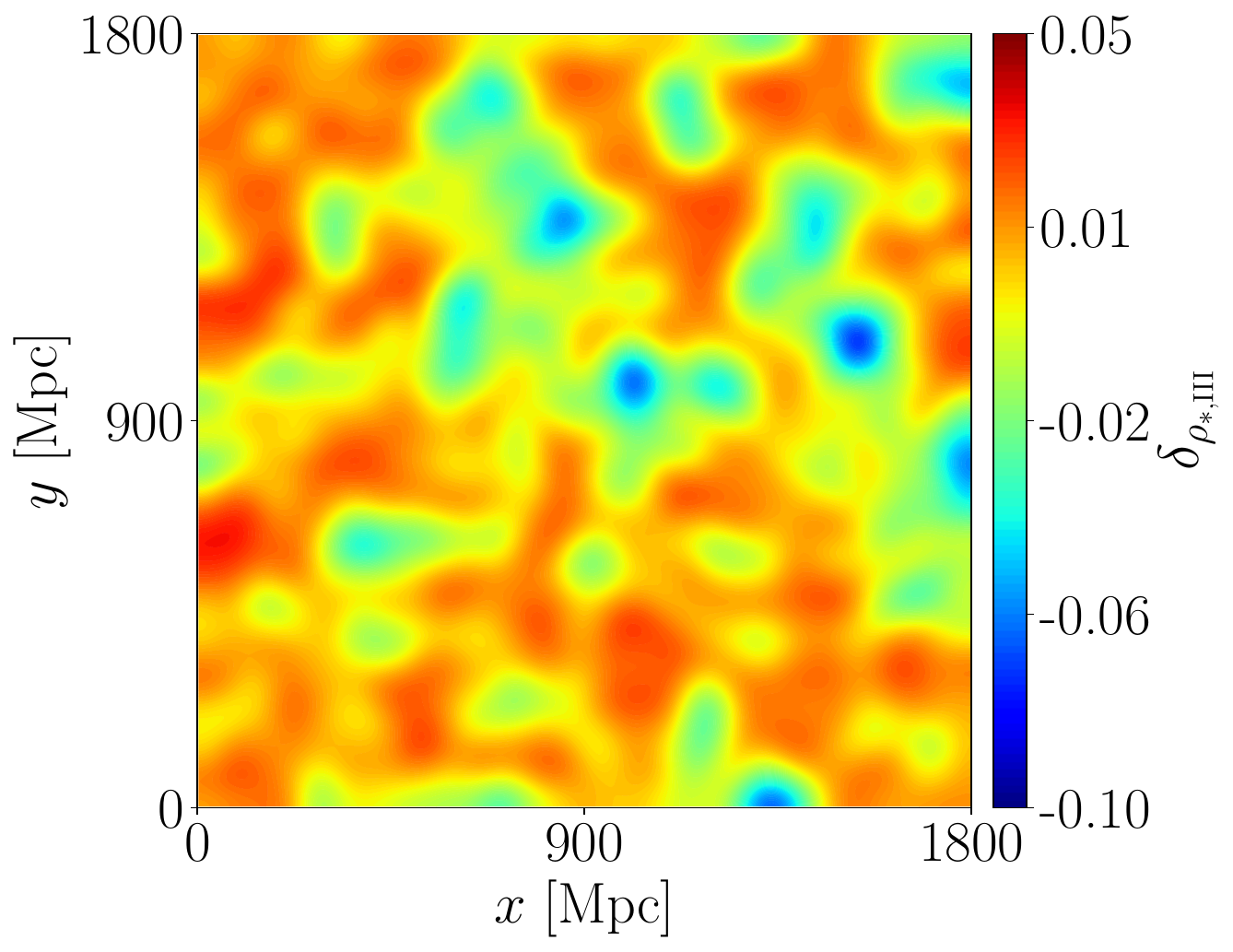}   
    \includegraphics[width=0.25\linewidth]{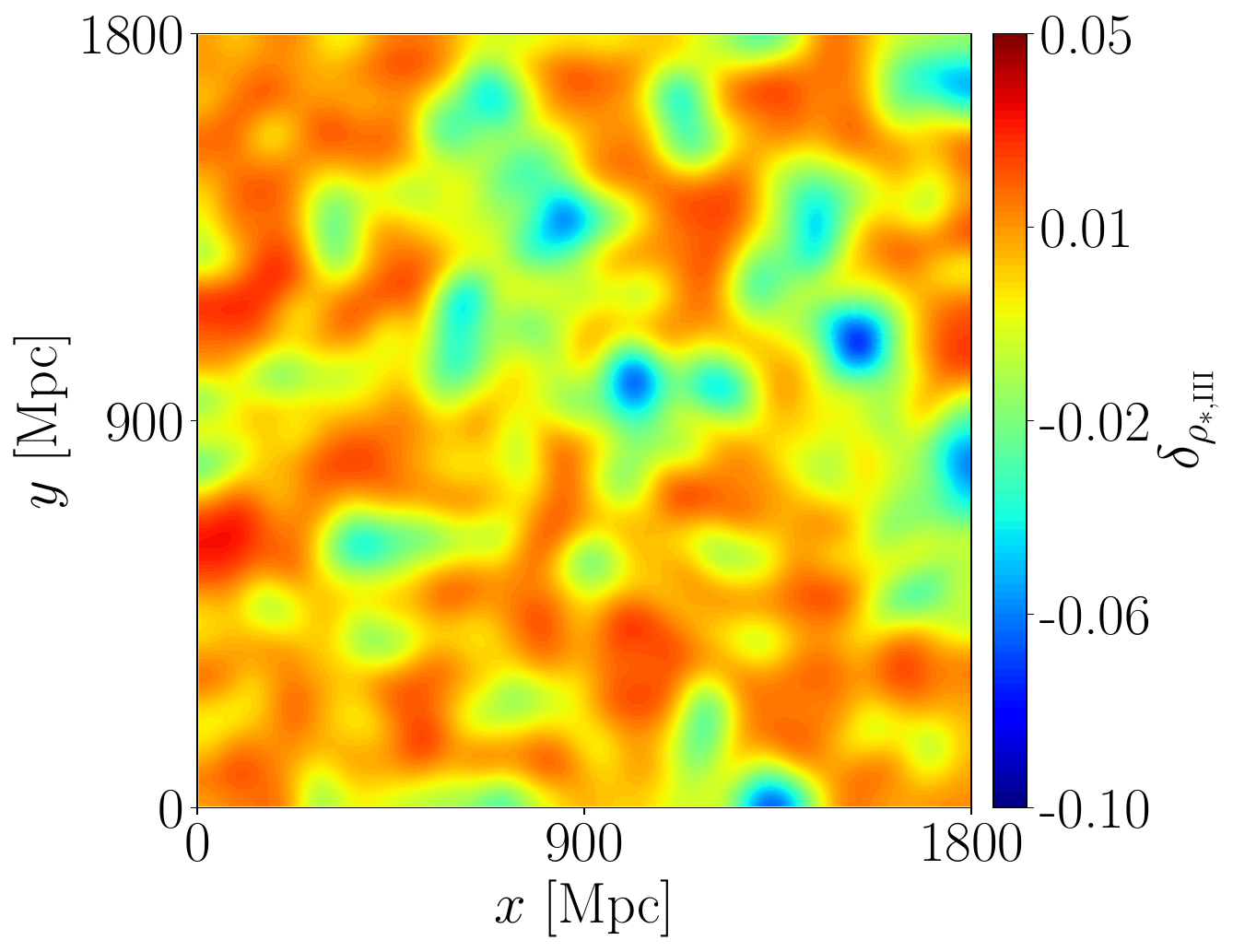}  
\includegraphics[width=0.25\linewidth]{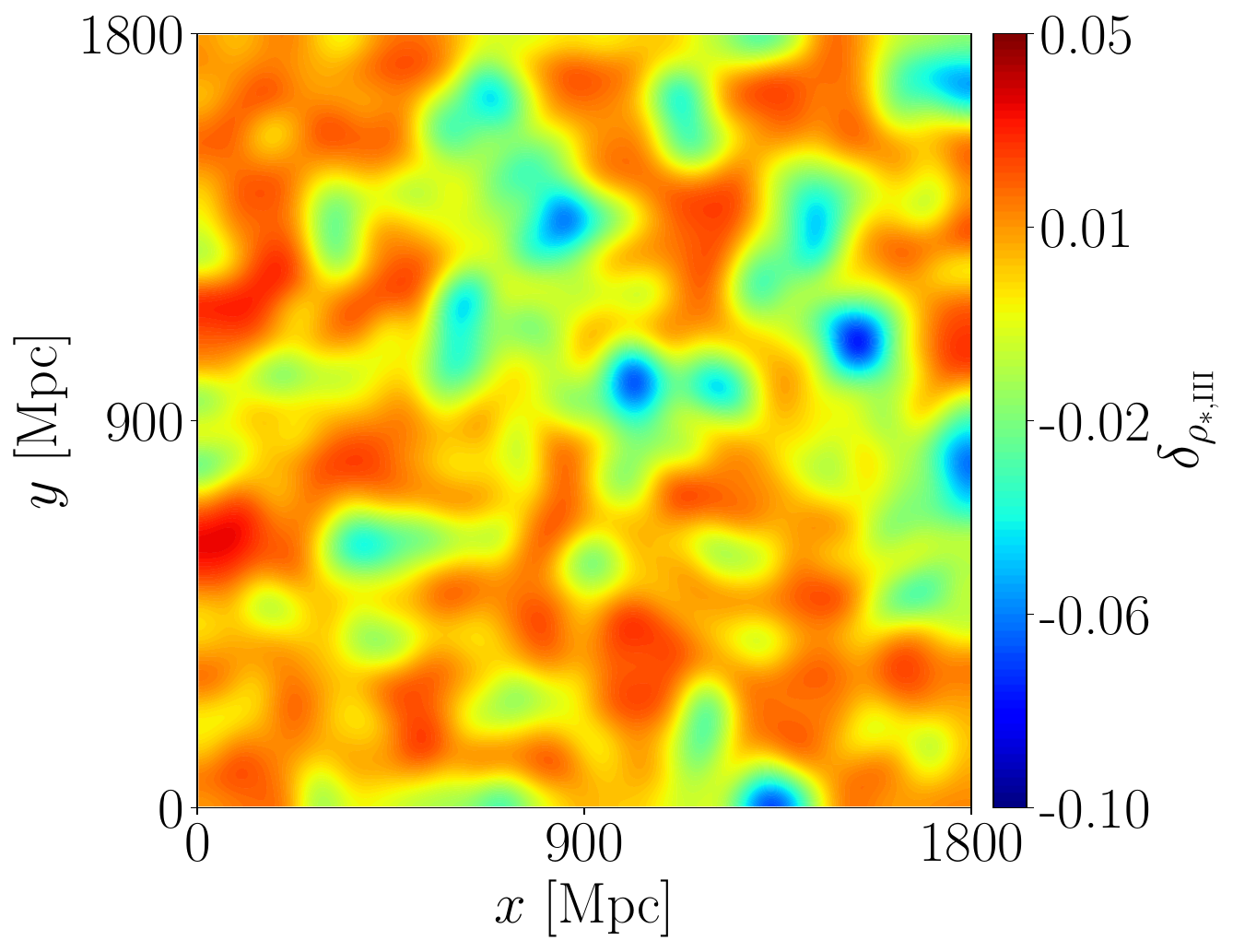}  
    \caption{The relative density contrast of the Pop III stars mass field, $\delta_{\rho_{*,\rm III}}=\rho_{*,\rm III}/\overline{\rho}_{*,\rm III}-1$. 
    In the {\it left} column, $M_{\rm cool0}=2.4\times 10^6~M_\odot$ and $\alpha_{\rm LW}=0$. In the {\it middle} column, $M_{\rm cool0}=10^5~M_\odot$, $\alpha_{\rm LW}=7$, and $J^{\rm V14}_{\rm LW}(z)=10^{-\frac{(z-20)}{10.0}}$. 
    In the {\it right} column $M_{\rm cool0}=10^5~M_\odot$, $\alpha_{\rm LW}=7$, and $J^{\rm V14}_{\rm LW}(z)=10^{-(z-20)}$.  
     {\it Top} row refers to $z=26$ while {\it bottom} $z=20$.
     At $z=20$, $M_{\rm cool1}$ is the same for all columns. To highlight the large-scale features, we smooth the field by a Gaussian window with radius 50 Mpc.
    }
    \label{fig:fcoll}
\end{figure*}

In Fig. \ref{fig:fcoll}  we plot the relative  density contrast field of the Pop III stars mass at $z=26$ and $z=20$, for the three illustrations.  The contrast is defined as $\delta_{\rho_{*,\rm III}}=\rho_{*,\rm III}/\overline{\rho}_{*,\rm III}-1$,  modulated by the $v_{\rm db}(\vec{r})$, where
\begin{equation}
\rho_{*,\rm III}(\vec{r}) \propto \int_{M_{\rm cool2}(\vec{r})} ^{ M_{\rm T_4}} M_{\rm h} \frac{dn}{dM_{\rm h}} dM_{\rm h},
\end{equation}
for which $M_{\rm T_4}$ is the virial mass corresponding to virial temperature of $10^4$ K, $dn/dM_{\rm h}$ is the halo mass function. Since here is an intuitive estimation, to focus on the influence of the relative streaming motion, we assume the density field is uniform, $dn/dM_{\rm h}$ is the Universal mass function \citep{shethEllipsoidalCollapseImproved2001}. Because the VAO features are large-scale features, to highlight the large-scale distribution of Pop III stars, we smooth the field by Gaussian window with radius 50 Mpc.

Fig. \ref{fig:fcoll} clearly shows that the influence of $M_{\rm cool0}$ and $\alpha_{\rm LW}$ on the Pop III stars field depends on LW radiation evolution. Although the illustrations (1) and (2) have different $M_{\rm cool0}$ and $\alpha_{\rm LW}$, their $\delta_{\rho_{*,\rm III}}$ fields are quite similar at $z=26$ and 20. That is to say, when LW radiation evolves slowly, $\alpha_{\rm LW}$ is somewhat degenerate with $M_{\rm cool0}$. However, for the illustration (3) the $\delta_{\rho_{*,\rm III}}$ field is obviously different from the illustration (1) at $z=25$. It implies that, if the living time of VAO features is longer than the typical evolution timescale of LW radiation, then it breaks the degeneracy between $M_{\rm cool0}$ and $\alpha_{\rm LW}$. Such degeneracy actually reflects the evolution of LW background.

\begin{figure}
    \centering
        \includegraphics[width=\linewidth]{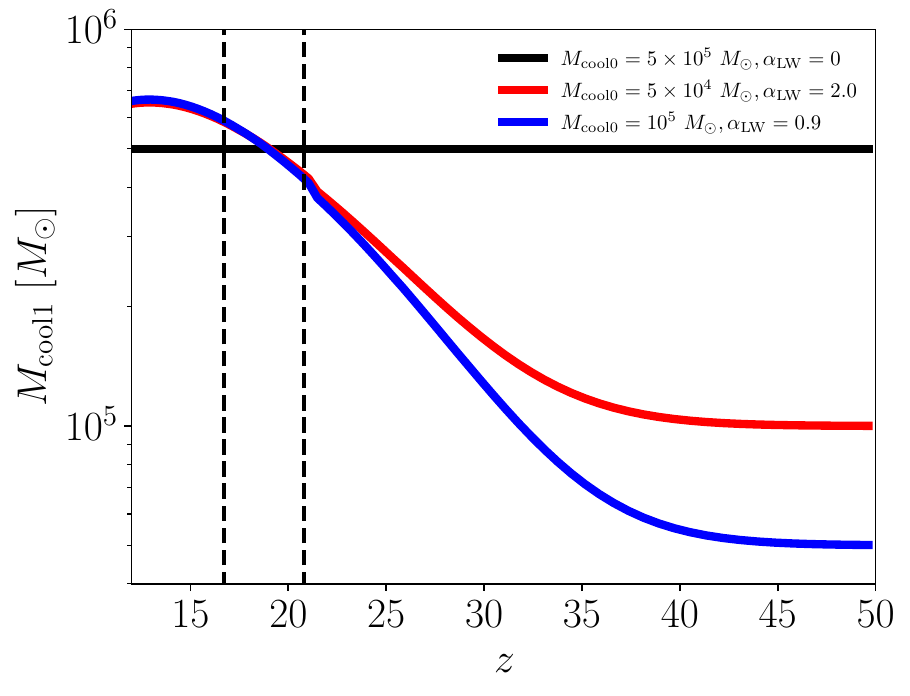}
    \caption{The $M_{\rm cool1}$ as a function of $z$, in three models with $M_{\rm cool0}=5\times 10^5~M_\odot$, $\alpha_{\rm LW}=0$; $M_{\rm cool0}=10^5~M_\odot$ and $\alpha_{\rm LW}=0.9$; and $M_{\rm cool0}=5\times 10^4~M_\odot$ and $\alpha_{\rm LW}=2.0$. We mark the range where VAO features appear, from $z\sim 21$ to $\sim 17$, by two vertical dashed lines.
    }
    \label{fig:Mcool1}
\end{figure}

We then perform simulations using the methods of \citet{Zhang_2024}, except that the critical mass for Pop III stars formation is replaced with $M_{\rm cool2}$ in Eq. (\ref{eq:Mcool2}). To deepen the understanding of degeneracy, before analyzing the 21 cm signal, in Fig. \ref{fig:Mcool1} we show the $M_{\rm cool1}$ as a function of $z$ for three models with: $M_{\rm cool0}=5\times 10^5~M_\odot$, $\alpha_{\rm LW}=0$; $M_{\rm cool0}=10^5~M_\odot$ and $\alpha_{\rm LW}=0.9$; and $M_{\rm cool0}=5\times 10^4~M_\odot$ and $\alpha_{\rm LW}=2.0$. In these three models, $M_{\rm cool1}$ is the same at $z\sim 19$, and the VAO features are visible from $z\sim 21$ to $\sim 17$. For the model with $5\times 10^4~M_\odot$ and $M_{\rm cool0}=10^5~M_\odot$, the VAO features appear slightly earlier than $M_{\rm cool0}=5\times 10^5~M_\odot$, consistent with the trend of $M_{\rm cool1}$. However the difference is very small, therefore these three models are actually in degeneracy.

\subsubsection{The mock 21 cm lightcone and multi-frequency angular power spectrum samples}

Instead of using the coeval boxes, here we must construct the lightcone from the series of output boxes. Because of this lightcone effect, the VAO features only occupy a narrow slice along the line of sight compared with the full length of  the coeval boxes. This allows us to generate many lightcone realizations by adding random  translations, rotations and reflections to the boxes. For each simulation, we generate 36 such realizations. Each realization is a mock observational sample, and their discrepancy represents the cosmic variance.

\begin{figure}
    \centering
    \includegraphics[width=0.9\linewidth]
{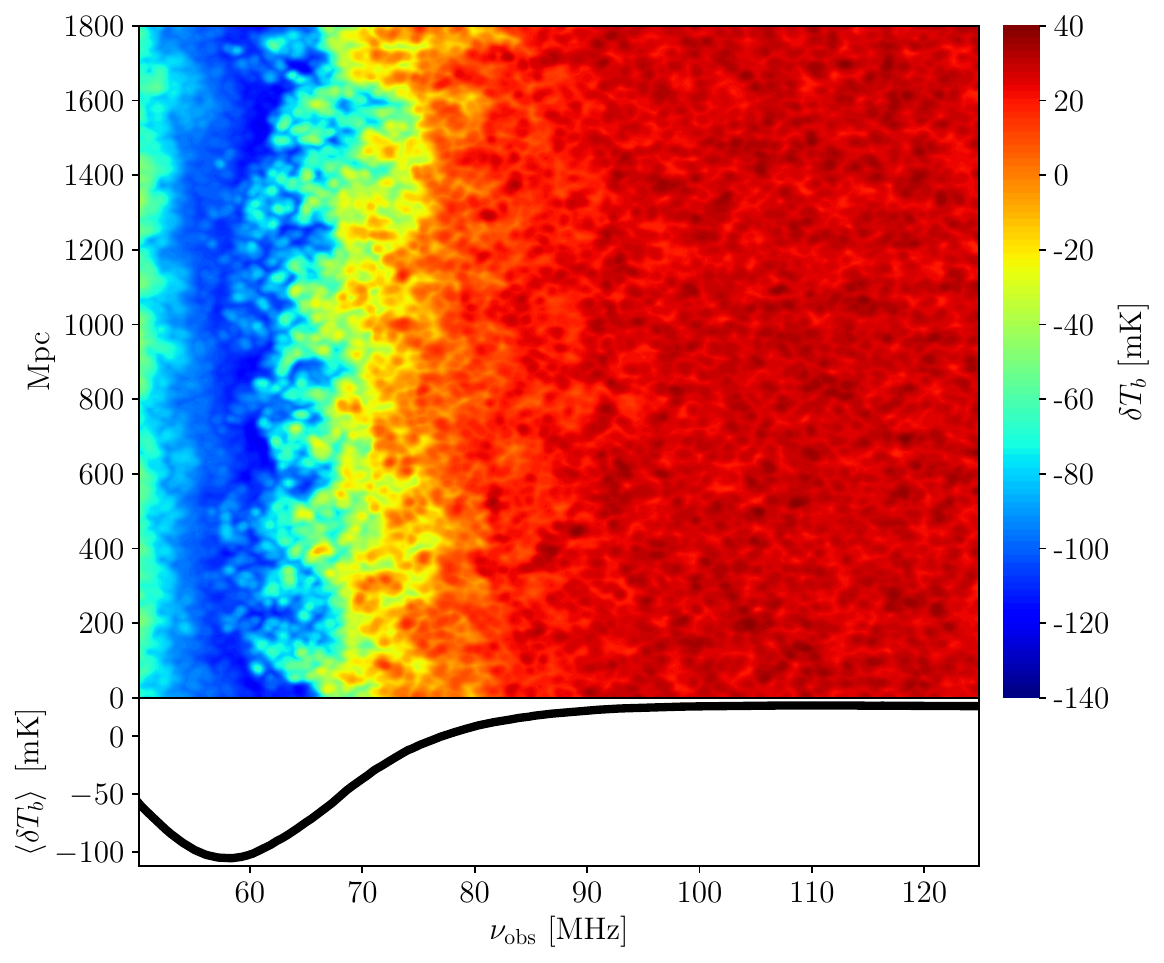}
    \caption{
   {\it Top:} A slice of the 21 cm lightcone, for $M_{\rm cool0}=10^5~M_\odot$ and $\alpha_{\rm LW}=0$.
   {\it Bottom:} The 21 cm global spectrum corresponding to the lightcone.
    }
    \label{fig:evolutionMap}
\end{figure}

In Fig. \ref{fig:evolutionMap} we show a slice of a lightcone constructed from the series of output coeval boxes in a simulation with $M_{\rm cool0}=10^5$ and $\alpha_{\rm LW}=0$ . The lightcone has redshift range from $z\sim 10$ to $z\sim 27$ (frequency from 50 - 125 MHz), and the transverse length 1800 Mpc, corresponding to a field of view $\sim 10.7^\circ$. This is a bit larger than the primary beam of a SKA-low station, whose diameter is 38 m \citep{dewdneySKA1DesignBaseline2022}. However, SKA-low allows to divide the station into multiple substations with diameter 18 m or even 12 m \citep{trottSKALowSubstationTemplates2024}, therefore feasible for observing such large-scale features.

\setcounter{footnote}{0}

There are different power spectrum estimators for the 21 cm spatial fluctuations: the three-dimensional power spectrum $P_{21}(\boldsymbol{k})$, the spherically-averaged three-dimensional power spectrum $P_{21}(k)$, the cylinder power spectrum $P_{21}(k_\perp,k_\parallel)$, and the angular power spectrum $P(k_\perp)$, where $k_\perp$ ($k_\parallel$) is the wavenumber perpendicular to (along) the line of sight \footnote{Because the locations of the VAO wiggles are constant in  $k_\perp$ space, here we use the power spectrum obtained by direct Fourier transform of the two-dimensional signal slice as the angular power spectrum. Its relation with the more popular form, $C(l)$, is 
\begin{align}
l&=k_\perp \times r(z) \nonumber \\
C(l)&=r^{-2}(z)P(k_\perp) \nonumber,
\end{align}
where $r(z)$ is the comoving distance of the source.}.

In Cosmic Dawn, the 21 cm signal evolves rapidly (e.g. \citealt{gharaMorphologyRedshifted21cm2024,cohenEmulatingGlobal21cm2020,beraStudyingCosmicDawn2023}) and it induces the lightcone effect (e.g. \citealt{dattaLightconeEffectReionization2012,ghara21CmSignal2015,Greig2018MNRAS.477.3217G}). Because of this effect, the $P_{21}(k)$ calculated from coeval boxes, like in \citet{Zhang_2024}, may overestimate the signal and cannot fairly count the contributions from directions along and perpendicular to the line of sight.  \citet{munozImpactFirstGalaxies2022a} compared the VAO signal from different slices in the lightcone, found that the effect can reduce the signal by $\sim 20\%$.

In this paper we use the MAPS \citealt{Santos2005ApJ...625..575S,Datta2007MNRAS.378..119D,mondalPredictionsMeasuring21cm2020,mondalMultifrequencyAngularPower2022,Trott2022A&A...666A.106T,Shaw2023MNRAS.522.2188S}), $P_{21}(k_\perp,\nu_{\rm obs})$, as the estimator of the VAO features. Compared with the spherically-averaged three-dimensional power spectrum, the MAPS does not need to perform Fourier transform along line of sight, therefore suitable for analyzing the rapidly-evolved signal.

\begin{figure}
    \centering
    \includegraphics[width=0.9\linewidth]
{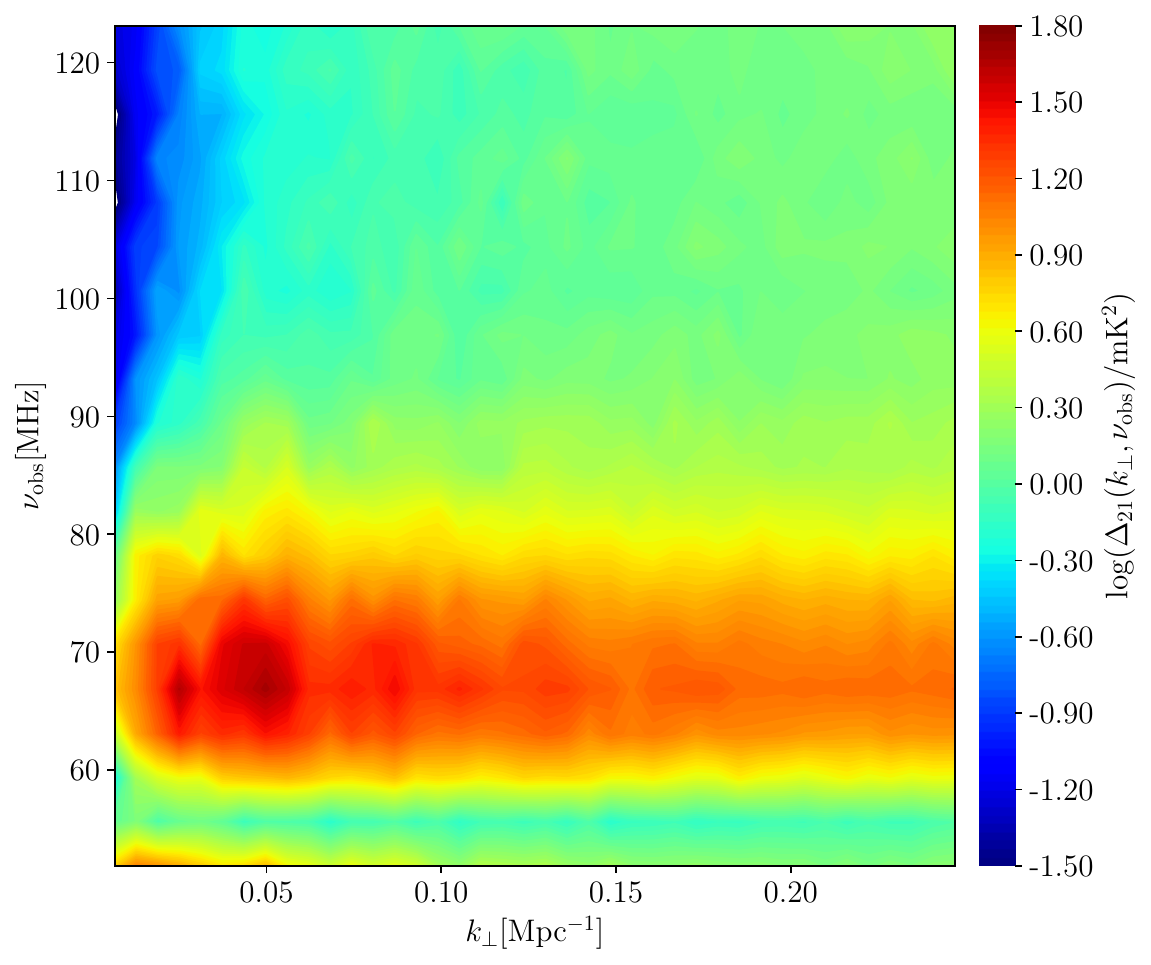}
    \caption{The MAPS map calculated from the lightcone of Fig. \ref{fig:evolutionMap}.    
    }
    \label{fig:angularPS}
\end{figure}

Fig. \ref{fig:angularPS} shows one MAPS map realization calculated from the lightcone in Fig. \ref{fig:evolutionMap}. Each angular power spectrum has frequency width $\Delta \nu_{\rm obs}=3.75$ MHz. It evolves rapidly, and exhibits clearly the VAO wiggles. The wiggles start to appear at $\nu_{\rm obs}\sim 60$ MHz, reach  maximum at $\nu_{\rm obs}\sim 70$ MHz, and disappear after  $\nu_{\rm obs}\sim 80$ MHz. The maximum bandwidth for the exhibition of the wiggles is just $\sim 20$ MHz, corresponding to a comoving length $\sim 500\ \mathrm{Mpc}$ along the line of sight.

The angular power spectrum can be decomposed into two parts: a smooth component $P_{\rm 21,smooth}(k_\perp,\nu_{\rm obs})$ and a component of net wiggles $P_{\rm 21,wiggles}(k_\perp,\nu_{\rm obs})$, and generally 
\begin{equation}
|P_{21,\rm wiggles}(k_\perp,\nu_{\rm obs})|\lesssim 10\% |P_{\rm 21, smooth}(k_\perp,\nu_{\rm obs})|.
\end{equation}
Note however, the relative streaming motion not only produces the wiggles component, but may also changes the shape and amplitude of the smooth component \citep{Visbal2012Natur.487...70V,munozStandardRulerCosmic2019,sarkarMeasuringCosmicExpansion2023}. 
For the following reasons, in this paper we only use the net wiggles component to constrain the baseline cooling threshold parameter and the LW feedback efficient parameter: (1) In principle, if the entire power spectrum is used to constrain parameters (e.g. \citealt{sarkarMeasuringCosmicExpansion2023}), the contribution of the VAO is naturally involved. However, in this case the constraints would be led by the signal irrelevant to the VAO, since in the entire power spectrum, the contribution from the VAO is minor, and there might be degeneracy. (2) It is easy to separate a net wiggles component and a smooth component, but difficult to distinguish between a smooth component relevant to the VAO and a smooth component irrelevant to the VAO, as they are both smooth. (3) Compared with the smooth component, the locations of the wiggles are purely determined by cosmology and fundamental physics \citep{munozStandardRulerCosmic2019,sarkarMeasuringCosmicExpansion2023},
and will not be confused with the noise and residual foreground. The amplitudes of the wiggles and their redshift evolution could be a good indicator for astrophysical effects.

To get the  net VAO wiggles, we fit the angular power spectrum $\Delta(k_\perp)=k^2_\perp/(2\pi)P(k_\perp)$ of each frequency bin by a 6th degree polynomial for $\ln k_\perp$ \citep{munozRobustVelocityinducedAcoustic2019},
\begin{equation}
\Delta^2_{21,\rm poly}(k_\perp,\nu_{\rm obs})=\sum_{i=0}^6c_i(\nu_{\rm obs})(\ln\textit{k}_\perp)^i,
\end{equation}
and then remove it. The residual 
\begin{equation}
\Delta^2_{21,\rm wiggles}(k_\perp,\nu_{\rm obs})=\Delta^2_{21}(k_\perp,\nu_{\rm obs})-\Delta^2_{21,\rm poly}(k_\perp,\nu_{\rm obs}),
\end{equation}
is the net VAO wiggles. We check that, using 4th to 7th degree polynomials, the net wiggles are almost identical. For 3th degree polynomial, the wiggles are slightly larger, while for 8th and 9th degree polynomials the wiggles are slightly lower. The detailed results and relevant discussion are shown in Sec. \ref{sec:degree_poly}.

\begin{figure*}
    \centering
    \includegraphics[width=0.33\linewidth]{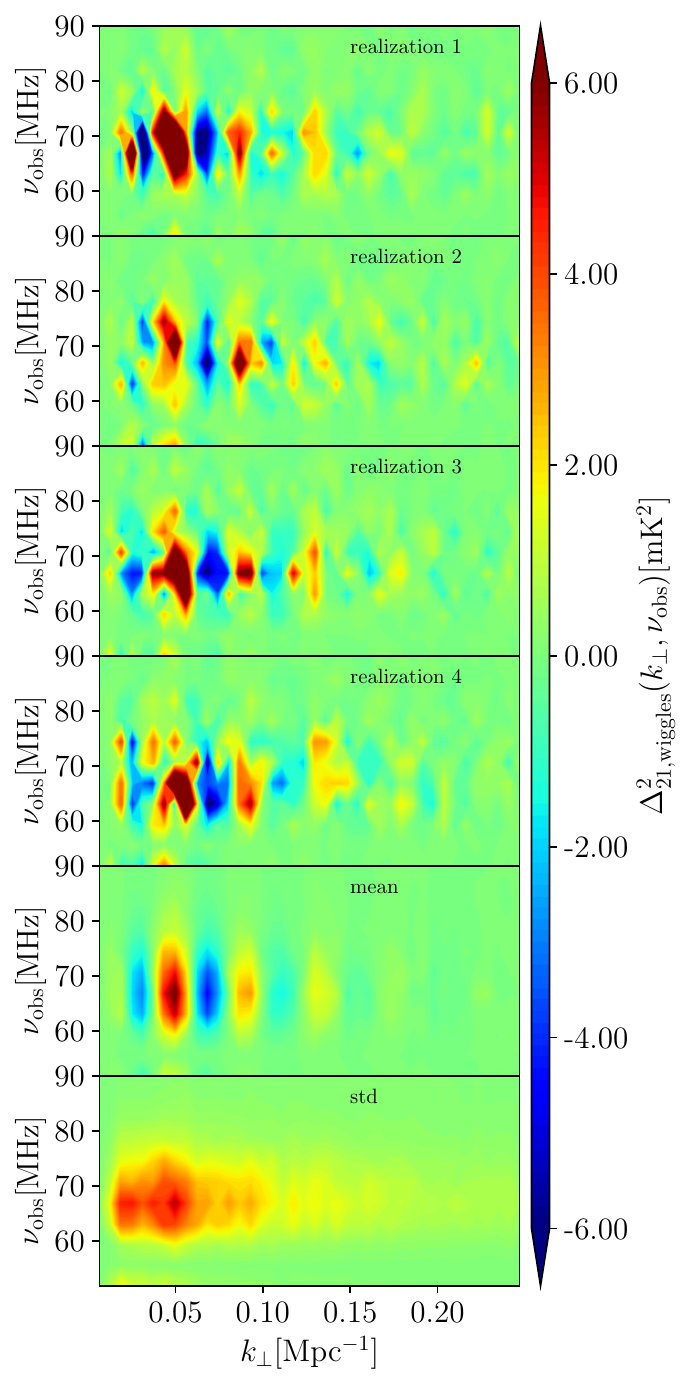}  
\includegraphics[width=0.33\linewidth]{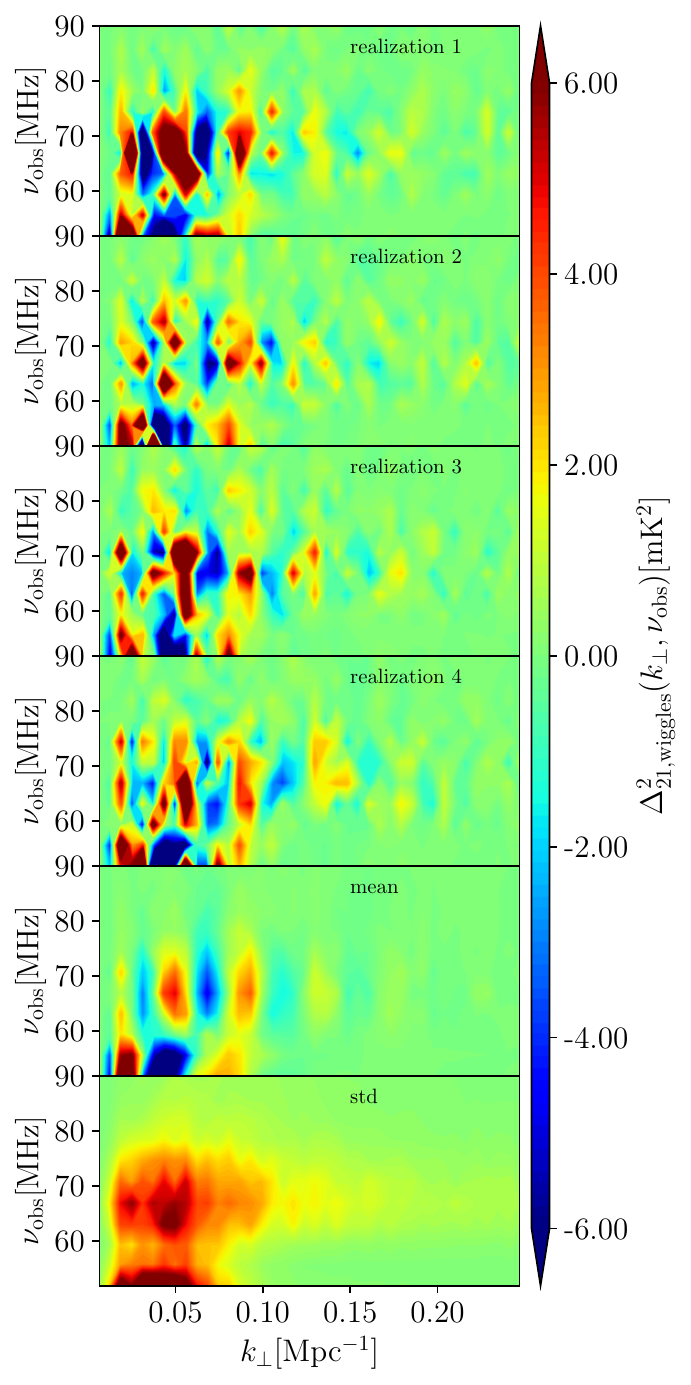}  
 \includegraphics[width=0.33\linewidth]{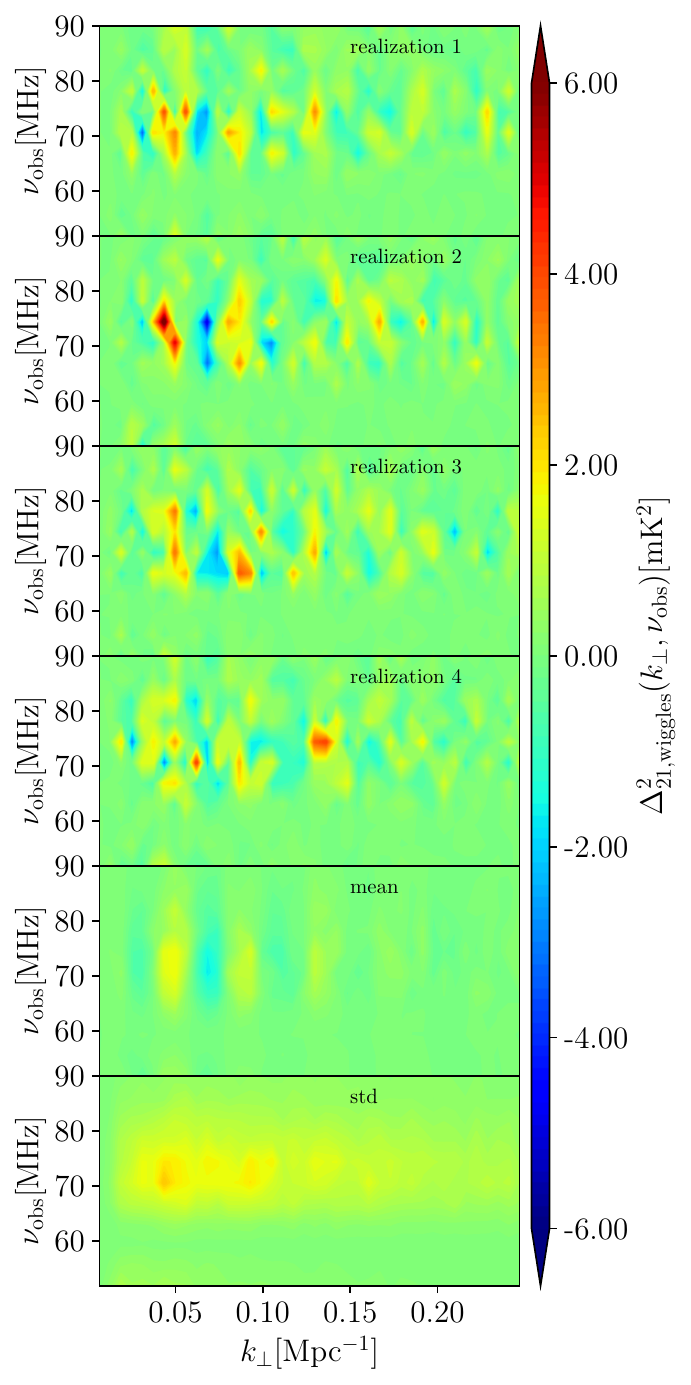}  
\includegraphics[width=0.33\linewidth]{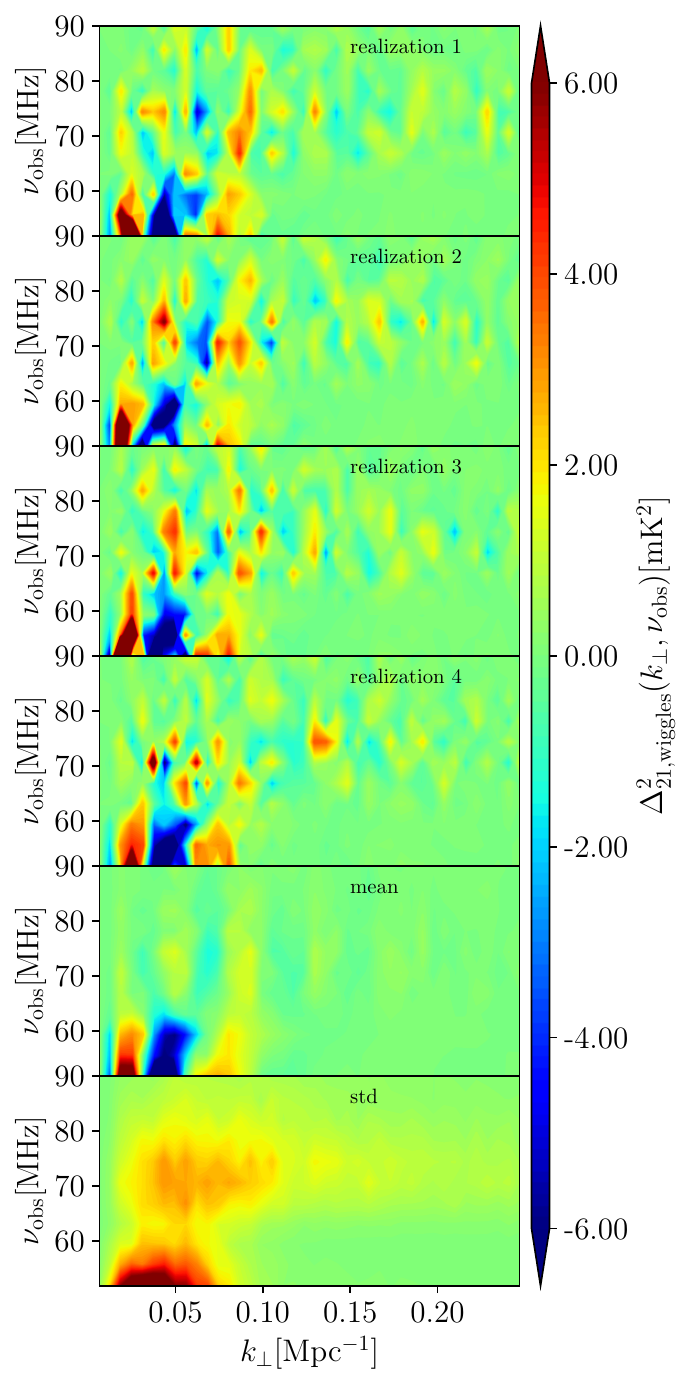}  
    \caption{
{\it Top lef column:} The VAO wiggles in the MAPS maps, for $M_{\rm cool0}=10^5~M_\odot$ and $\alpha_{\rm LW}=0$. {\it Top right column:} Similar to the first column but with the noise of the SKA-low AA* with $10^4$ hour integration time. {\it Bottom left column:} Similar to the first column, but for $M_{\rm cool0}=10^5~M_\odot$ and $\alpha_{\rm LW}=4.0$. {\it Bottom right column:} noise of the SKA-low AA* with $10^4$ integration hour is added to the third column. One simulation produces 36 lightcone realizations. For each column, we randomly select 4 realizations and show them in the top four panels. The fifth panel is the mean MAPS of the 36 realizations, while the last panel is the standard deviation of the 36 realizations.  For displaying purpose, we ignore the $\nu_{\rm obs} > 90$ MHz part, where the wiggles are invisible.
    }
    \label{fig:meanVAO}
\end{figure*}

For a simulation with $M_{\rm cool0}=10^5~M_\odot$ and $\alpha_{\rm LW}=0$, in the first column of Fig. \ref{fig:meanVAO} we show the net VAO wiggles for 4 realizations randomly selected from the 36 realizations produced by this simulation, and the mean and standard deviation of the 36 realizations.

Strong wiggles are obviously seen between $\nu_{\rm obs}\sim 60$ MHz and $\nu_{\rm obs}\sim80$ MHz. This is promising for further analysis as the signal is sufficiently strong. However, we also note that the variations of the wiggles for different realizations are rather large, it implies that the cosmic variance could be a significant source for uncertainties.
Although the VAO wiggles only exist from $\sim 60$ MHz to $\sim 80$ MHz, we use the MAPS map in the full frequency range $50 - 125$ MHz to derive the parameters. This is because when $M_{\rm cool0}$ and $\alpha_{\rm LW}$ change, the frequency range of the VAO wiggles may also change. But we check that, our results will not be influenced by the frequecy with empty VAO wiggles.

\begin{figure}
    \centering
    \includegraphics[width=1\linewidth]{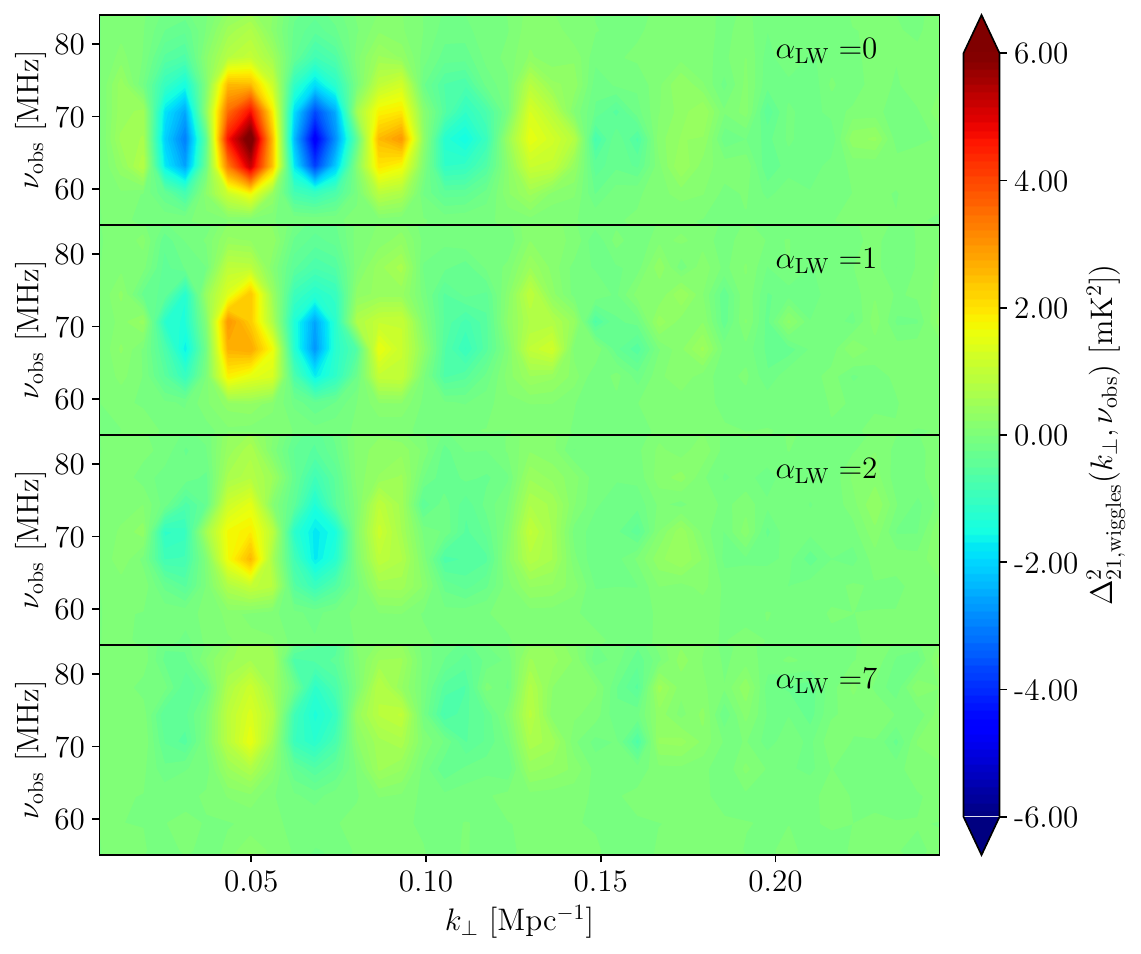} 
   \caption{The VAO features for $M_{\rm cool0}=10^5~M_\odot$ and various $\alpha_{\rm LW}$, as labeled in each panel. For displaying purpose, we only show the frequency range where the VAO features are visible.
    }
    \label{fig:VAO_alpha_LW}
\end{figure}

In Fig.  \ref{fig:VAO_alpha_LW} we plot the $\Delta_{21,\rm wiggles}(k_\perp, \nu_{\rm obs})$ for $M_{\rm cool0}=10^5~M_\odot$ however $\alpha_{\rm LW}$ changes from 0 to 7. Clearly, the VAO features monotonously decrease with increasing $\alpha_{\rm LW}$. Indeed, they can be good indicator for the LW feedback efficient.

\begin{figure*}
    \centering
    \includegraphics[width=0.9\linewidth]{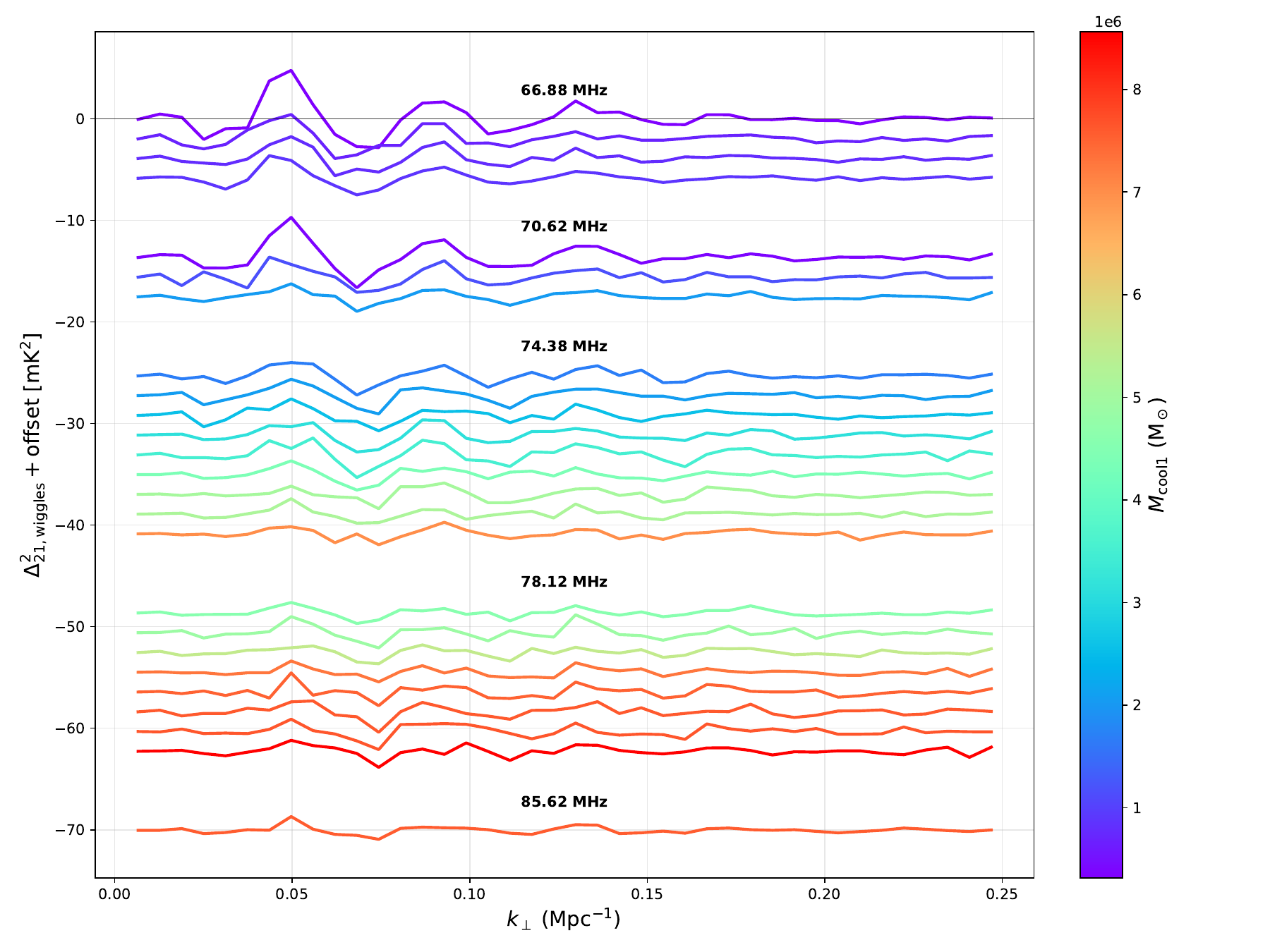}
    \caption{
    The VAO wiggles for various $M_{\rm cool0}$ and $\alpha_{\rm LW}$ values. For displaying purpose, we shift the curves according to the their frequency along the $y$-axis. The curve is colored by the $M_{\rm cool1}$ value.
    }
    \label{fig:VAO_variables}
\end{figure*}

In Fig. \ref{fig:VAO_variables} we plot the $\Delta_{\rm 21,wiggles}(k_\perp,\nu_{\rm obs})$ at the frequency where the first peak at $k_\perp=0.05$ Mpc$^{-1}$ reaches maximum, for various $M_{\rm cool0}$ and $\alpha_{\rm LW}$ parameters. Generally, the smaller the $M_{\rm cool1}$, the larger the wiggles. This figure also shows that different $M_{\rm cool1}$ may produce similar VAO features at similar redshift, implying that $\alpha_{\rm LW}$ would be in degenerate with $M_{\rm cool0}$.

\subsection{Machine Learning and Parameter Inference}

\begin{figure}
    \centering
    \includegraphics[width=0.9\linewidth]{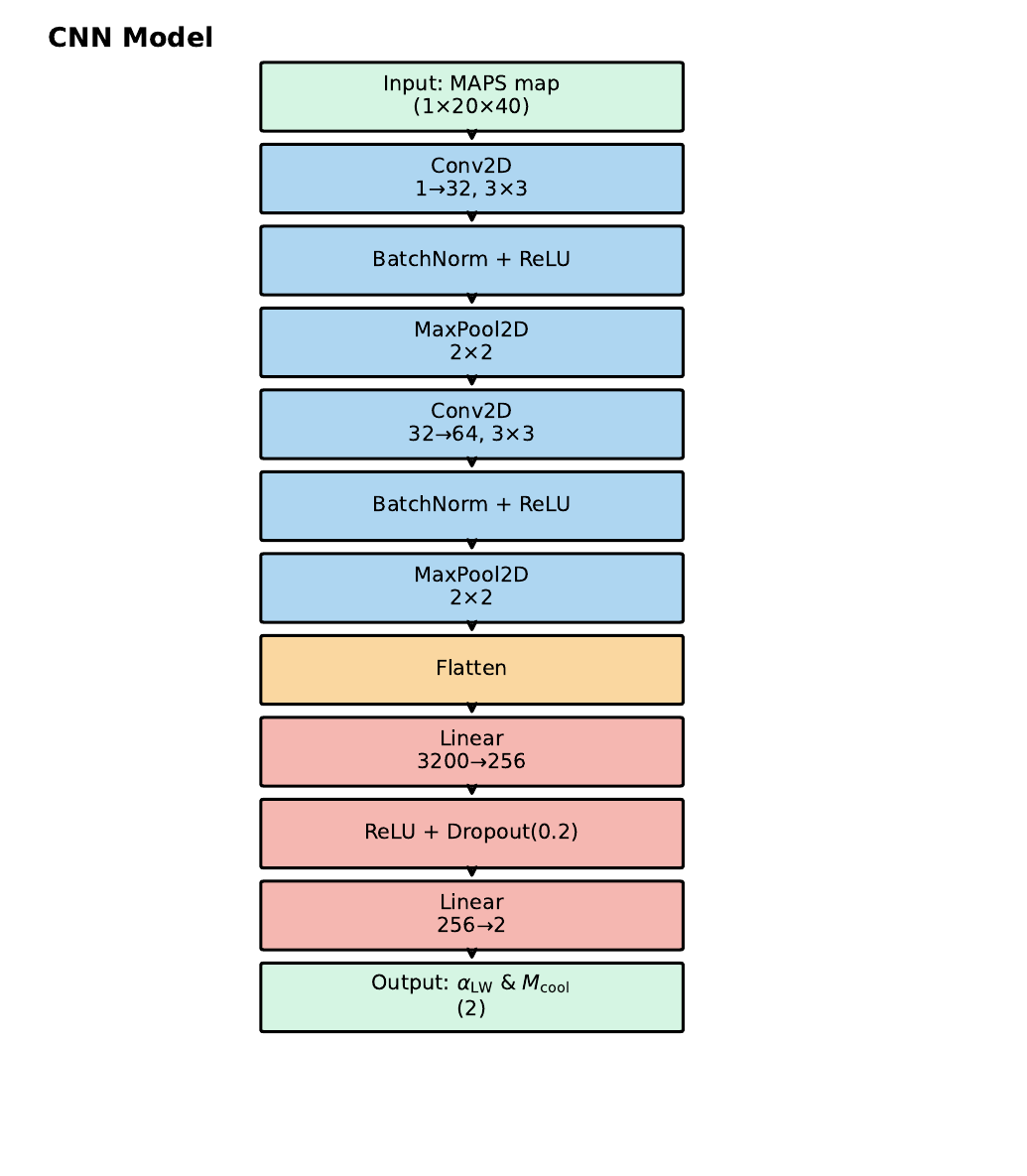}
    \caption{An illustration of our CNN model. We have two convolutional layers (Conv2D-BatchNorm-ReLU-MaxPool) , a flatten layer and two fully connected layers (Linear).
     }
    \label{fig:flowchart}
\end{figure}

The relations between the parameters in Eq. (\ref{eq:Mcool1}) and the MAPS maps are rather complicated, and can only be built by simulations. We employ the machine learning (ML) technique for parameter inference. It is more efficient in our case. Moreover,  compared with the traditional MCMC algorithm, the ML method has the advantage that it does not need to assume the explicit likelihood function, therefore is applicable for non-Gaussian likelihood (\citealt{sunDeepLearningdrivenLikelihoodfree2025,2025arXiv251010713T,2025arXiv250922561M,2025arXiv250208152B}). 
We adopt the Convolutional Neural Network (CNN) for processing the data in the ML method. Our neural network consists of two convolutional layers and two fully connected layers.  We make a flowchart to show how it works, see Fig. \ref{fig:flowchart}. We implement an early stopping mechanism during training to prevent overfitting.

The parameter inference process performed by the CNN on our MAPS maps is essentially point estimation, meaning that we cannot obtain the posterior distribution of the parameter estimates. This issue can certainly be addressed by using the Bayesian Neural Networks like in \citet{Zhao_2022}. 
However, in our work, each simulation with the same $M_{\rm cool0}$ and $\alpha_{\rm LW}$ can produce many random MAPS map realizations with different cosmic variance or noise variance. The distribution of the recovered parameters naturally gives the parameter uncertainties led by these  variances.

We require the CNN to estimate $M_{\rm cool0}$ and $\alpha_{\rm LW}$ simultaneously, however, their values differ by orders of magnitudes. We therefore replace them with  normalized variables when train the CNN model. Take $\alpha_{\rm LW}$ as example, for which the corresponding normalized variable is 
$y^{\alpha_{\rm LW}} = (\alpha_{\rm LW} - \bar{\alpha}_{\rm LW}) / \sigma_{\alpha_{\rm LW}},$
where $\bar{\alpha}_{\rm LW}$ is the mean $\alpha_{\rm LW}$ of all training samples, and $\sigma_{\alpha_{\rm LW}}$ is their standard deviation. The same applies to $M_{\rm cool0}$.

We use the loss function (mean squared error, MSE), mean absolute error (MAE) and the standard recovery performance $R^2$ to estimate the goodness of the parameter inference. For a normalized parameter the loss function is actually  the relative error. The recovery performance indicates the overall parameter estimation capability of the model. The closer $R^2$ is to 1, the better the model's parameter estimation. Note that $R^2$ is not necessarily to be positive, however if it is much smaller than 1, generally it means the recovery is bad. Moreover, $R^2$ depends on the intrinsic scatter of the true values. If the sample has low diversity, even the predicted values are close to the true values, the recovery performance is still interpreted as not good.

\section{Results} \label{sec:results}

In this section, we investigate three cases: inferring the LW feedback efficient parameter $\alpha_{\rm LW}$, inferring $\alpha_{\rm LW}$ and the baseline cooling threshold parameter $M_{\rm cool0}$ simultaneously in the absence/presence of noise. Their characteristics are summarized  in Tab. \ref{tab:cases}.

\subsection{Case A}

In this case, we run 500 simulations with 
$M_{\rm cool0}=10^5~M_\odot$ but $\alpha_{\rm LW}$ randomly distributed between 0.0 and 7.0. Each simulation produces 36 lightcone and MAPS map realizations. A MAPS map realization is a sample. It is an two-dimensional image with $20\times 40$ pixels, composed of 20 $k_\perp$ bins and 40 $\nu_{\rm obs}$ bins. We have 18000 samples in total, they are grouped into different sets.
The training set comprises 14400 samples, while the validation set and test set both contains 1800 samples.

\begin{figure}
    \centering
    \includegraphics[width=1\linewidth]{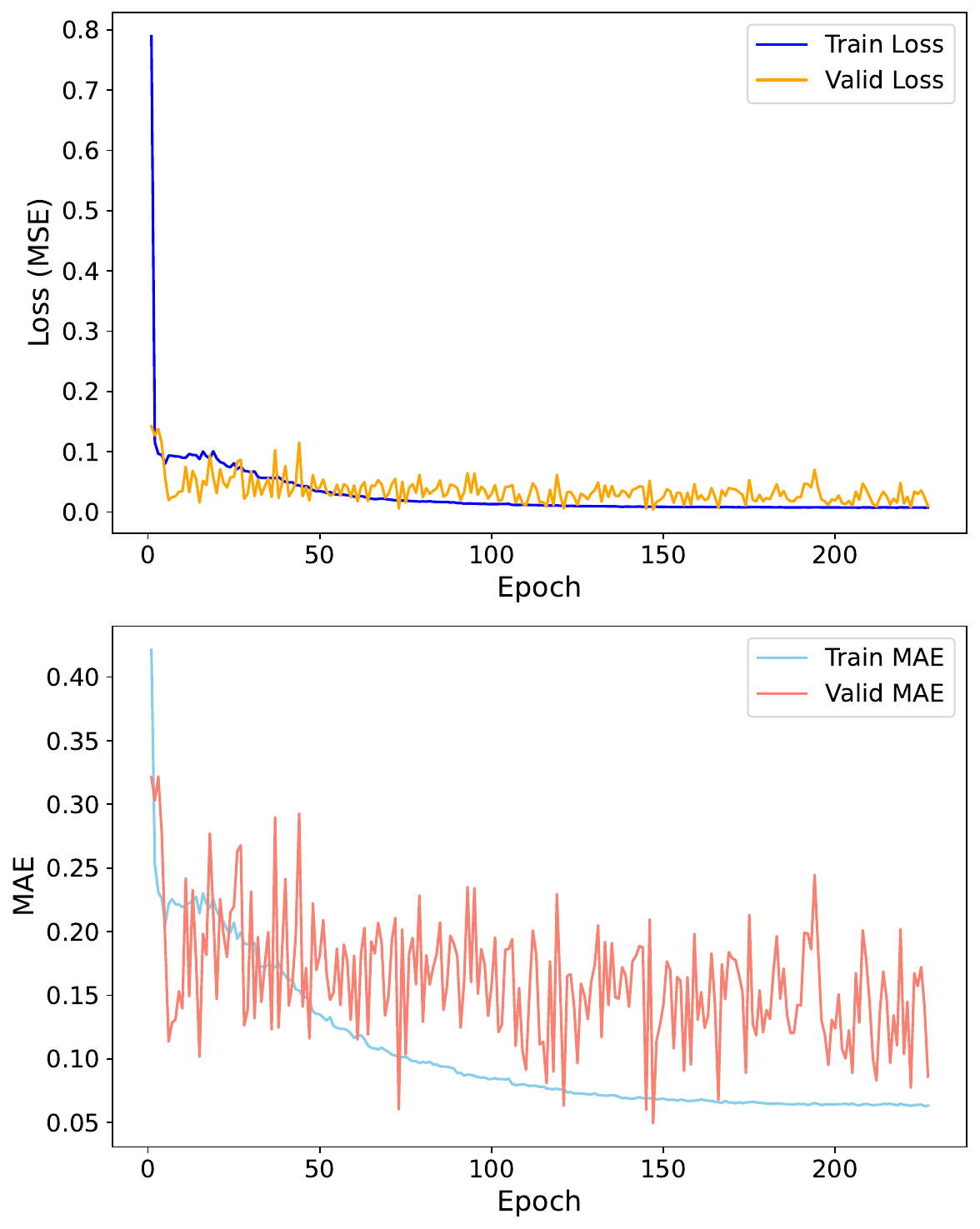}
    \caption{The training and validation curves of Case A. \textit{Top:} Loss function (MSE). \textit{Bottom:} MAE. }
    \label{fig:traininghistory_CaseA}
\end{figure}

In Fig. \ref{fig:traininghistory_CaseA} we show the training history, including the evolution of the loss function and the MAE in the training set and the validation set. The loss function of the training set converges stably with limited overfitting.

\begin{figure}
    \centering
    \includegraphics[width=1\linewidth]{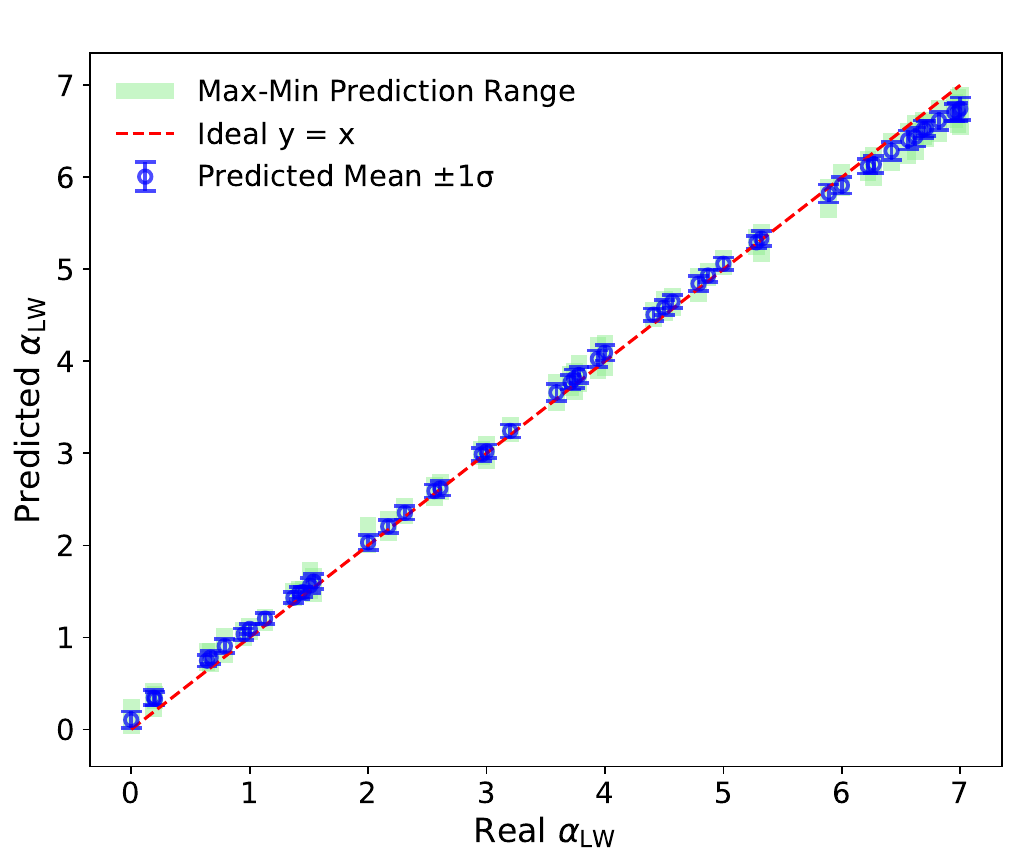} 
   \caption{The recovered $\alpha_{\rm LW}$ against the true values for Case A. 
   Filled patches are minimum and maximum ranges of the recovered parameters for same true values, while the points and errorars are mean and standard deviation of them. To guide the eye, we plot the $y=x$ relation by a dashed line.   
    }
    \label{fig:caseA}
\end{figure}

In Fig. \ref{fig:caseA} we show the recovered $\alpha_{\rm LW}$ against the true values. Such LW feedback efficient parameter is recovered very well, with loss function $L_{\alpha_{\rm LW}}=0.0045$ and recovery performance $R^2_{\alpha_{\rm LW}}=0.9961$. Since each simulation produces 36 lightcone and MAPS map realizations, the recovered $\alpha_{\rm LW}$ scatters around the true value. This actually represents the influence of cosmic variance. 

From Fig. \ref{fig:meanVAO}, we see that the VAO features have large cosmic variance, the observed sample for a $\sim 10.7^\circ$ FoV (represented by a single realization) could obviously deviate from the signal (represented by the mean MAPS). However, even so, in Fig. \ref{fig:caseA} $\alpha_{\rm LW}$ is still recovered very well.

\subsection{Case B}

The baseline cooling threshold parameter $M_{\rm cool0}$ is independent of the LW feedback and in principle could be derived from other independent observations, so that we can focus on the LW feedback efficient. However, VAO features also depend on this parameter and its uncertainties will finally influence the recovery of $\alpha_{\rm LW}$. To investigate this, we also run simulations with varying $M_{\rm cool0}$ and $\alpha_{\rm LW}$, and recover them simultaneously. 

We run 1000 simulations with $M_{\rm cool0}$ uniformly distributed between $5\times 10^4~M_\odot$ and $5\times 10^5~M_\odot$, and $\alpha_{\rm LW}$ uniformly distributed between 0.0 and 7.0. Again, each simulation produces 36 lightcone and MAPS map realizations. So we finally have 36000 samples, 28800 of them are used for training, 3600 are used for validation and 3600 are used as test set. 
 
\begin{figure}
    \centering
    \includegraphics[width=1\linewidth]{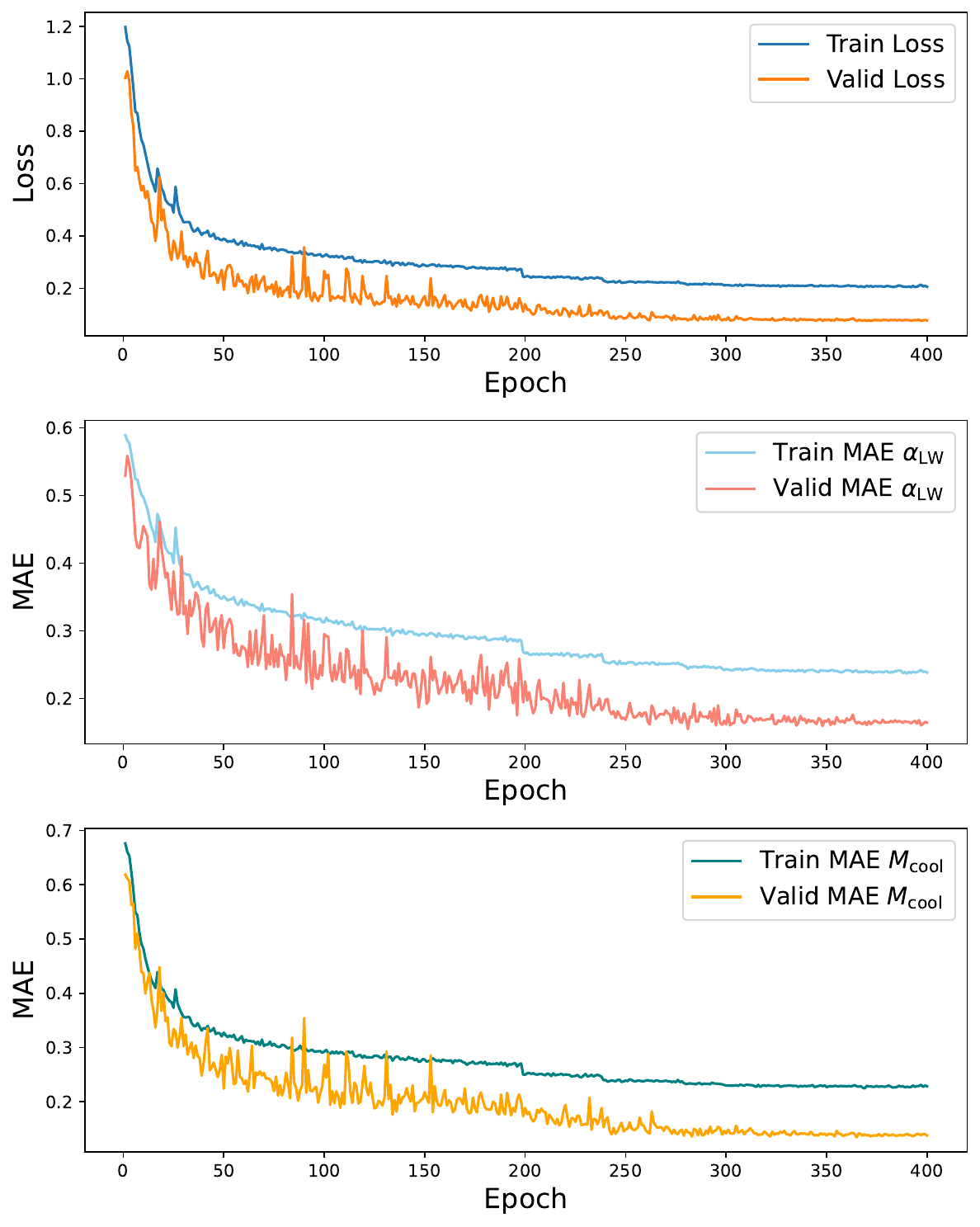}
    \caption{The training and validation curves of Case B. \textit{Top:} The loss function (MSE). \textit{Middle:} The MAE for $\alpha_{\rm LW}$ prediction. \textit{Bottom:} The MAE for $M_{\rm cool0}$ prediction.
    }
    \label{fig:traininghistory_CaseB}
\end{figure}

In Fig. \ref{fig:traininghistory_CaseB} we show the training history for Case B, including the evolution of loss function and MAE in training set and validation set respectively. The validation loss is lower than the training loss, because we adopted dropout and batch normalization that introduce stochasticity during training, while validation set is performed using the full deterministic model. 
After $\sim$300 learning epochs, both the loss and the MAE converged.

\begin{figure}
    \centering
    \begin{subfigure}{0.9\linewidth}
        \centering
        \includegraphics[width=\linewidth]{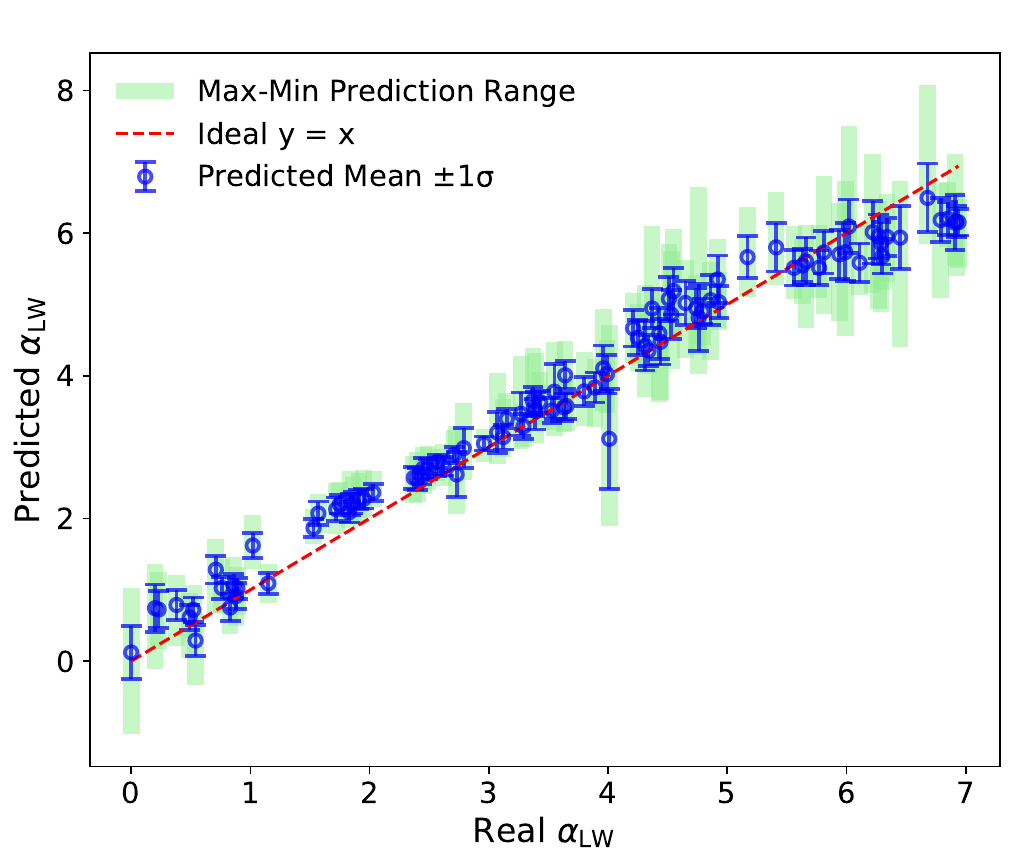}
    \end{subfigure}
    \hfill
    \begin{subfigure}{0.9\linewidth}
        \centering
        \includegraphics[width=\linewidth]{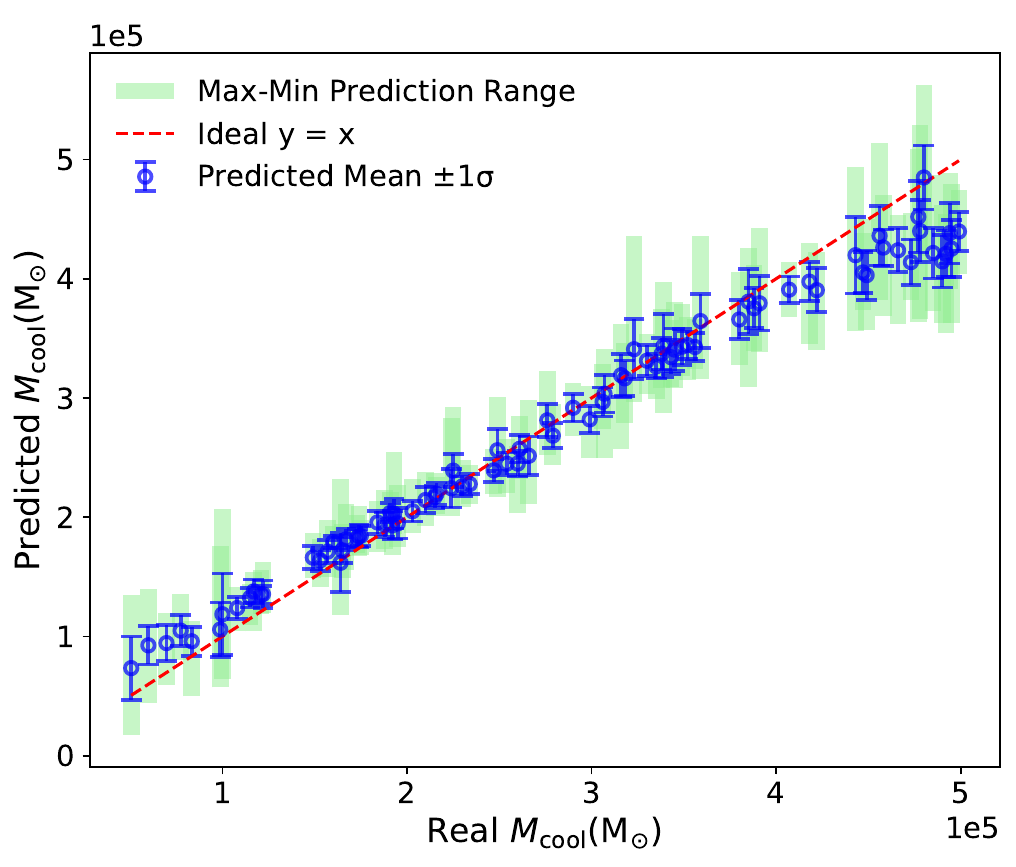}
    \end{subfigure}
    \caption{
    {\it Top:} The recovered $\alpha_{\rm LW}$ against the true values. 
    {\it Bottom:} Same to the top panel, however for the parameter $M_{\rm cool0}$. In all panels, filled patches are minimum and maximum ranges of the recovered parameters for same true values, points and errorbars are mean and standard deviation of them. The dashed curve is the $y=x$ relation for the purpose of guiding the eye.
    }
    \label{fig:errorCaseB}
\end{figure}

In Fig.\ref{fig:errorCaseB} we show the results of the recovered $\alpha_{\rm LW}$ and $M_{\rm cool0}$ against the true values. The total loss function is the sum of loss functions for them.  $L_{\rm tot}=L_{\alpha_{\rm LW}}+L_{M_{\rm cool0}}=0.0984$. The recovery performances $R_{\alpha_\mathrm{LW}}^2 = 0.9497$ and $R^2_{M_{\rm cool0}}=0.9470$.

For each $\alpha_{\rm LW}$ there are many MAPS maps with same or different $M_{\rm cool0}$ values, the recovered $\alpha_{\rm LW}$ values scatter around the true $\alpha_{\rm LW}$. The same applies to $M_{\rm cool0}$. In Fig. \ref{fig:errorCaseB}, the errorbars are the standard deviation of the recovered parameters for the same true values, regardless of another parameter. It is actually the distribution of the considering parameter after marginalizing another parameter. We remind that the uncertainties in Fig. \ref{fig:errorCaseB} include both the cosmic variance and the degeneracy between  $\alpha_{\rm LW}$ and $M_{\rm cool0}$. However, even in this case the recovery still has good performance.

\begin{figure}
    \centering
        \includegraphics[width=\linewidth]{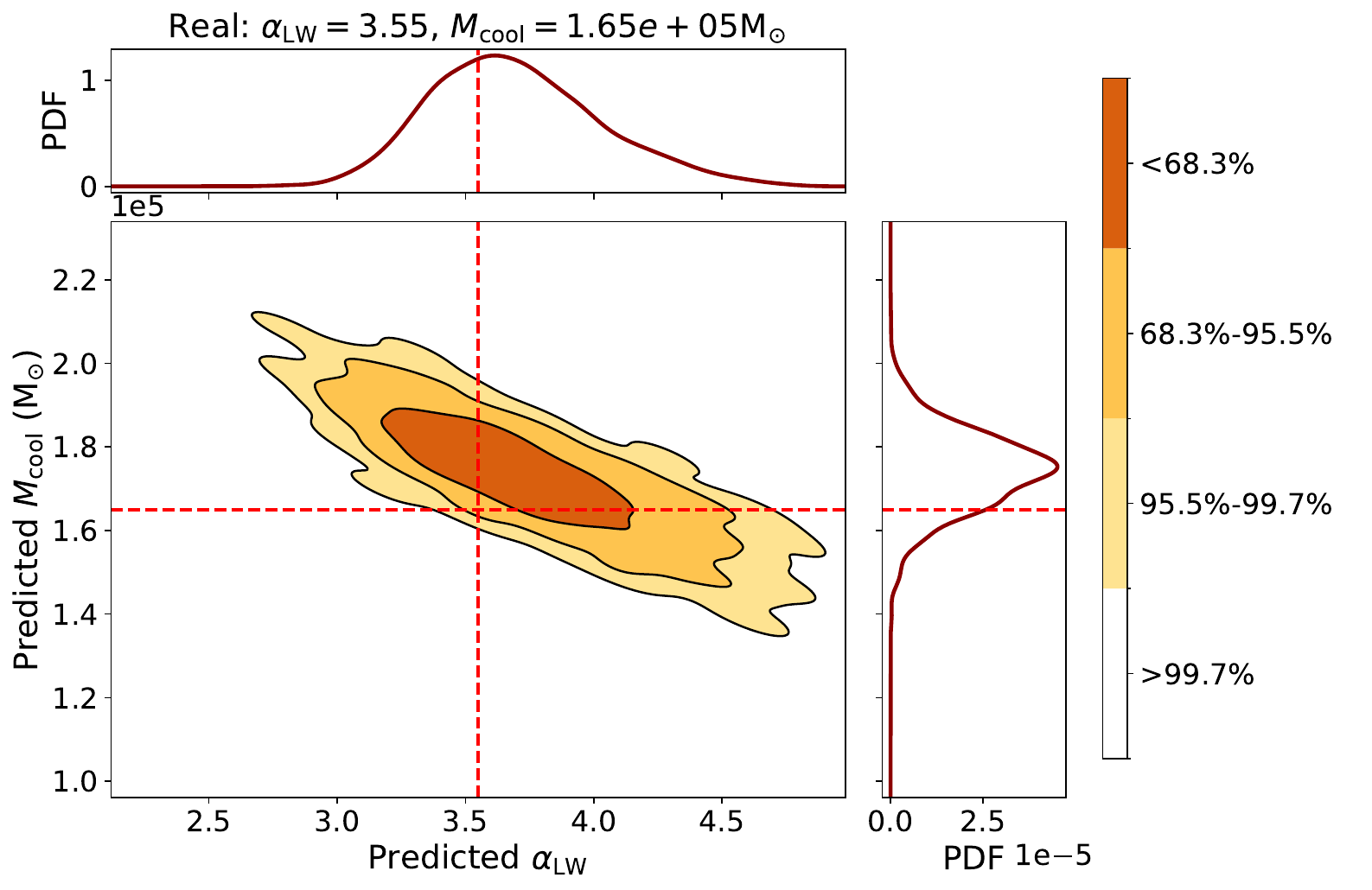}
    \caption{The joint probability density distribution of the predicted $M_{\rm cool0}$ and $\alpha_{\rm LW}$ for Case B. Dashed lines mark the input true values.
    }
    \label{fig:joint_PD}
\end{figure}

In Fig. \ref{fig:joint_PD} we show the joint probability density distribution of the predicted parameters in the $\alpha_{\rm LW}-M_{\rm cool0}$ panel. It shows clearly that the predicted $M_{\rm cool0}$ is correlated to the predicted $\alpha_{\rm LW}$. Smaller $M_{\rm cool0}$ and larger $\alpha_{\rm LW}$ produce similar VAO features with larger $M_{\rm cool0}$ and smaller $\alpha_{\rm LW}$. This is because in our model the real LW radiation evolves slowly in the redshift range where VAO features appear.
In a word: in Eq. (\ref{eq:Mcool1}), the degeneracy between the $M_{\rm cool0}$ and $\alpha_{\rm LW}$ do exist, the dependence of $J_{\rm LW}$ on $\vec{r}$ and $z$ helps to reduce some degeneracy,  but cannot fully remove it.

\subsection{Case C}

In order to estimate the influence of noise, we  generate MAPS samples with noise. The noise level is provided by SKA Sensitivity Calculator\footnote{\url{https://sensitivity-calculator.skao.int/}}, we adopt the SKA-low AA* (inner 350m) configuration, and set integration time $t_{\rm obs}=10^4$ hour. If the available survey time is 6 hours per day, then it takes 5 years. We assume the noise is Gaussian, and added it to the 21 cm lightcone, then use the same procedures to generate MAPS samples. The noisy MAPS maps are shown in right columns of Fig. \ref{fig:meanVAO}, compared with the samples without noise in left columns. The noise level rises dramatically below $\sim 60$ MHz. For $M_{\rm cool0}=10^5~M_\odot$, when $\alpha_{\rm LW}=0.0$, the VAO wiggles are strong enough, so the even in the presence of noise the features are still cearly visible. However when $\alpha_{\rm LW}=4.0$, the VAO wiggles become very weak, as a result when noise is involved, the features are hard to identify.

To improve the stability of training, for the mock sample with noise we replace the standard loss function with a Smooth L1 Loss, which is defined as
\begin{equation}
    L = 
    \begin{cases}
    \frac{1}{2}(y_\mathrm{pred} - y_\mathrm{true})^2 & |y_\mathrm{pred} - y_\mathrm{true}| < \beta \\
    \beta(|y_\mathrm{pred} - y_\mathrm{true}| - \frac{1}{2} \beta) & |y_\mathrm{pred} - y_\mathrm{true}| \geq \beta,
    \end{cases}
\end{equation}
 where $\beta$ is the connecting point of quadric and linear loss, and we set it 0.1. Such loss can perform better for noisy samples.

\begin{figure}
    \centering
    \includegraphics[width=0.9\linewidth]{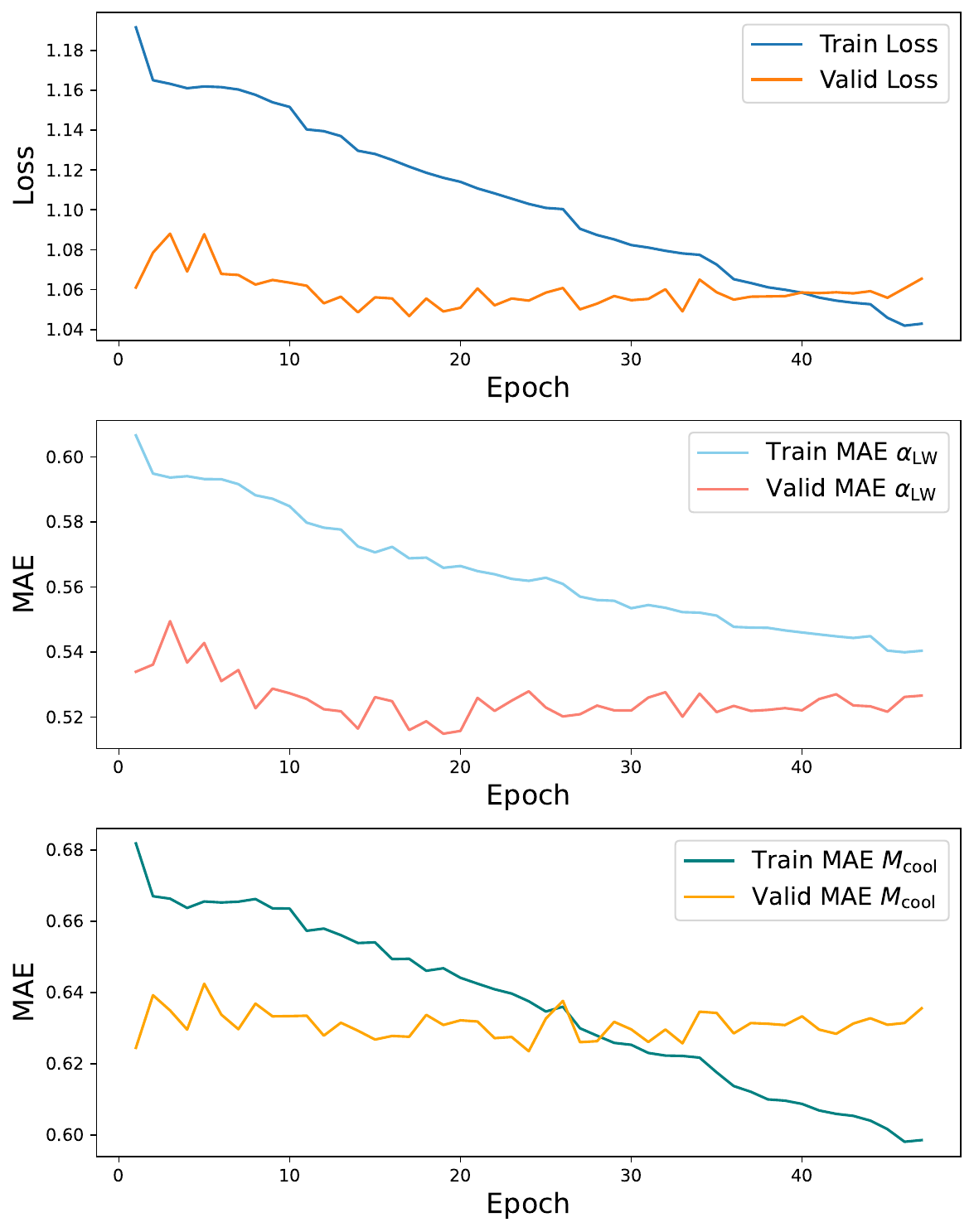}
    \caption{Training and validation curves of Case C. \textit{Top:} Loss function (MSE). \textit{Middle:} MAE of normalized $\alpha_{\rm LW}$ prediction. \textit{Bottom:} MAE of normalized $M_{\rm cool0}$ prediction.}
    \label{fig:traininghistory_CaseC}
\end{figure}

The evolution of the loss function and the MAE curves are shown in Fig. \ref{fig:traininghistory_CaseC}.  We find that when noise is involved, the loss function becomes much larger and the recovery performance becomes worse. Overfitting occurs at early epochs. 
We finally get $L_{\mathrm{tot}} = 1.0185$, $R_{\alpha_{\rm LW}}^2 = 0.5355$ and $R_{M_{\rm cool0}}^2 = 0.4081$.
For  $\alpha_{\mathrm{LW}}$ the recovery performance is $\sim 0.5$, allowing to at least  roughly estimate the range of $\alpha_{\mathrm{LW}}$. For $M_{\mathrm{cool0}}$, however,  the recovery performance is a bit worse, and it is hard to estimate $M_{\mathrm{cool0}}$ validly from  the  noisy MAPS sample. It is not clear why the performance of $M_{\rm cool0}$ is worse than $\alpha_{\rm LW}$; probably since the effects of $\alpha_{\rm LW}$ rely on the LW evolution, it carries more information.

\begin{figure}
    \centering
    \begin{subfigure}{0.9\linewidth}
        \centering
        \includegraphics[width=\linewidth]{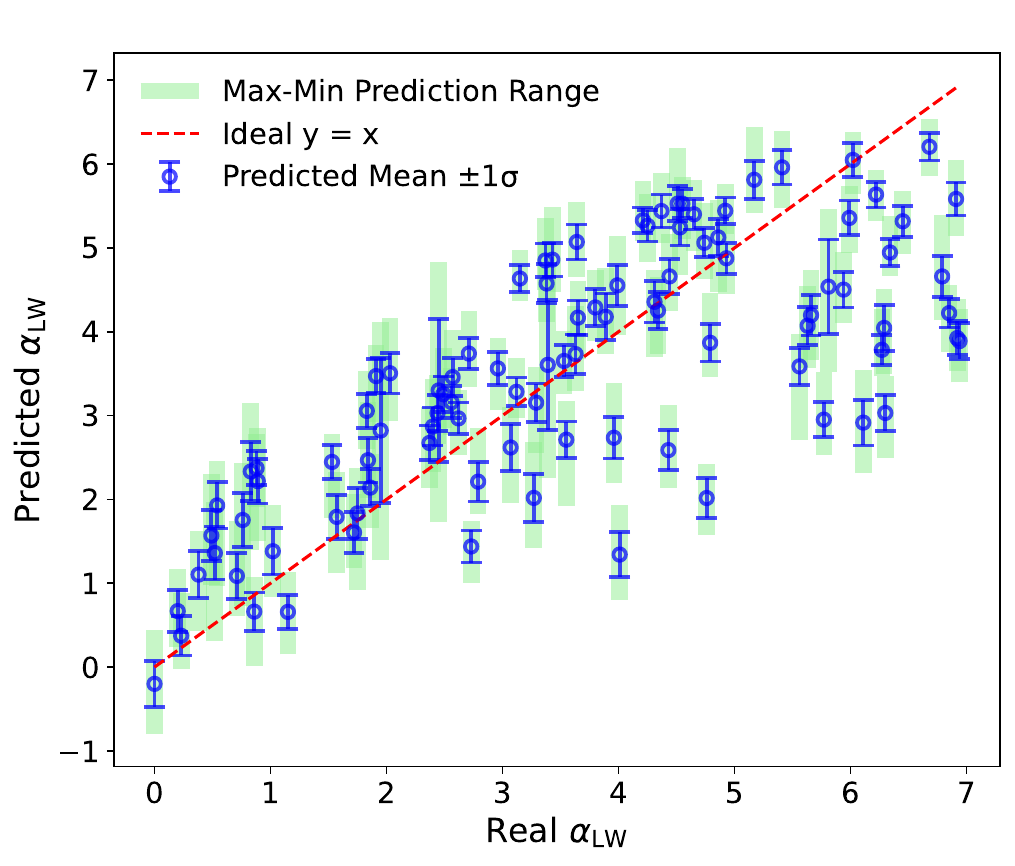}
    \end{subfigure}
    \hfill
    \begin{subfigure}{0.9\linewidth}
        \centering
        \includegraphics[width=\linewidth]{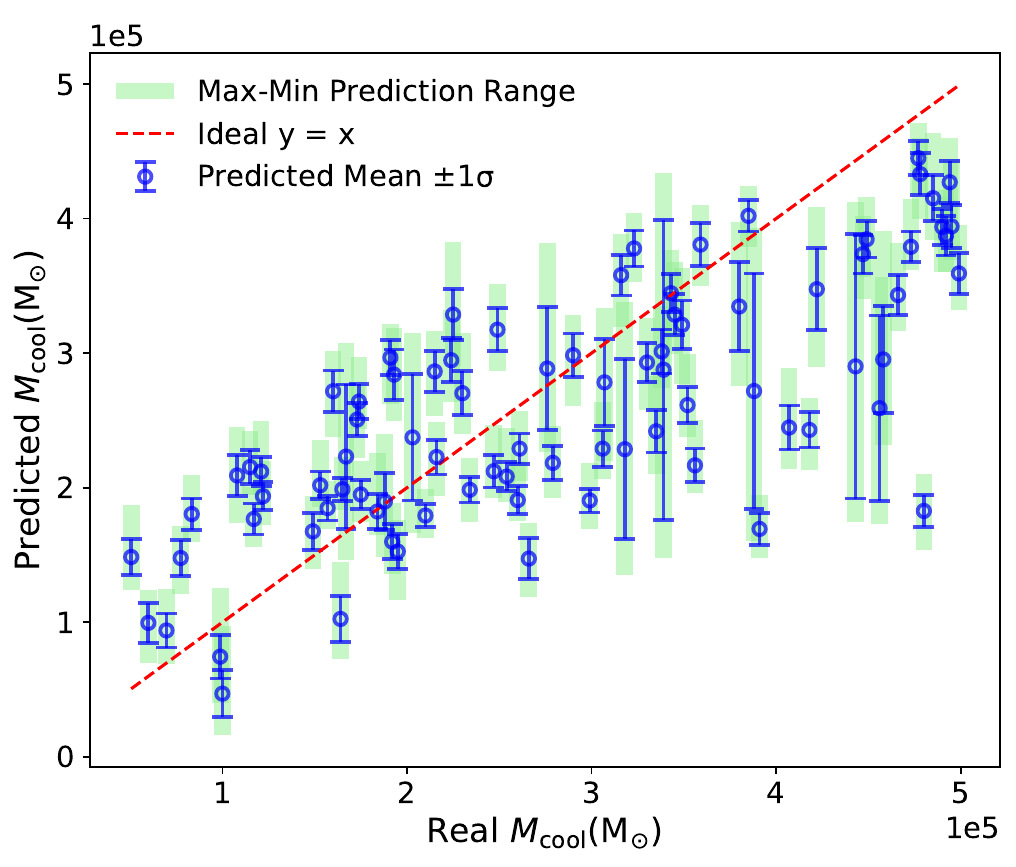}
    \end{subfigure}
    \caption{
    Similar to Fig.\ref{fig:errorCaseB}, however here the noise for SKA-low AA* is added to the mock samples.
    }
    \label{fig:errorCaseC}
\end{figure}

The recovered  $\alpha_{\rm LW}$ and $M_{\rm cool0}$ against the true values are plotted in Fig. \ref{fig:errorCaseC}. The uncertainties are much larger than the noise free case.

\begin{figure}
    \centering
    \begin{subfigure}{0.9\linewidth}
        \centering
        \includegraphics[width=\linewidth]{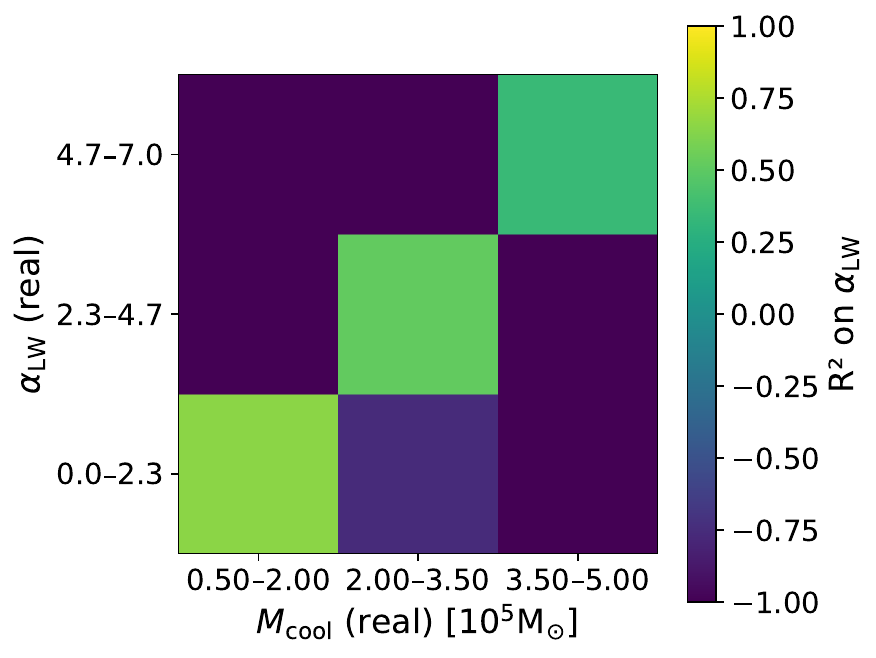}
    \end{subfigure}
    \hfill
    \begin{subfigure}{0.9\linewidth}
        \centering
        \includegraphics[width=\linewidth]{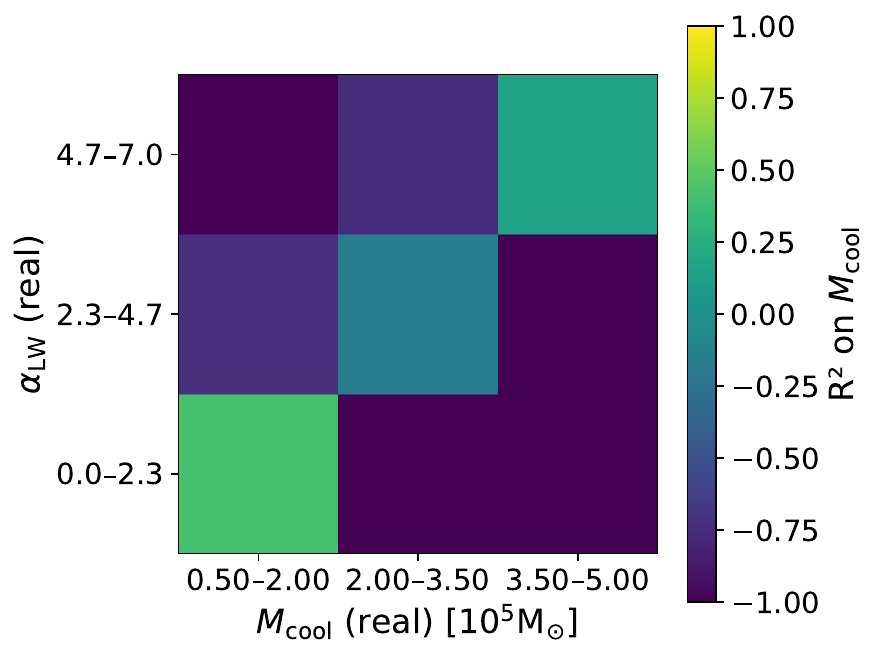}
    \end{subfigure}
    \caption{The local $R^2$ in $\alpha_{\rm LW}-M_{\rm cool0}$ plane, for case C. \textit{Top:} local $R_{\alpha_{\rm LW}}^2$. \textit{Bottom:} local $R_{M_{\rm cool0}}^2$.
    .
    }
    \label{fig:localCaseC}
\end{figure}

From Fig. \ref{fig:meanVAO}, we see that the influence of noise depends on the parameter values. To evaluate the recovery performance in different regions of the parameter space, we divide the $\alpha_{\rm LW}-M_{\rm cool0}$ plane into $3\times 3$ regions and show the local $R^2$ for each region in the plane, see Fig. \ref{fig:localCaseC}. It is natural that for the region with smallest $\alpha_{\rm LW}$ and $M_{\rm cool0}$, the signal is strongest and the recovery performance is the best. Moreover, we find that the recovery performance is better for the parameters in the diagonal. This is because in such two-parameter case, parameter degeneracy is a large uncertainty. Models with lower $M_{\rm cool0}$ and higher $\alpha_{\rm LW}$ may confuse the models with higher $M_{\rm cool0}$ and lower $\alpha_{\rm LW}$. Therefore the recovery performances for regions close to the top left and bottom right corners in the $\alpha_{\rm LW}-M_{\rm cool0}$ plane are worse.

Our results show that for $t_{\rm obs}=10^4$ hour, $\alpha_{\rm LW}$ and $M_{\rm cool0}$ are just marginally identified from the VAO features.
To recover them from  features in the MAPS sample, the required integration time for SKA-low AA* should be larger than $10^4$ hours. This is challenge but still feasible.

\begin{table*}
\caption{Characteristics of the three cases 
}
\hspace{0cm}
\begin{tabular}
{l|llllllllllllllllll}

&  {\bf Case A} & {\bf Case B} &{\bf Case C}  \\  
\hline
 Target Parameters &   $\alpha_{\rm LW}$ & $\alpha_{\rm LW}$ \& $M_{
 \rm cool0
 }$ & $\alpha_{\rm LW}$ \& $M_{
 \rm cool0
 }$   \\
  No. of Simulations &   500 &  1000 & 1000 \\
  No. of Samples &   18000 & 36000 &  36000  \\
No. of Training Samples &   14400 & 28800 &  28800  \\
No. of Validation Samples &   1800 & 3600 &  3600  \\
 No. of Test Samples &   1800 & 3600 &  3600  \\  
  
Noise &    No&  No &  Yes\\  
$L_{\rm tot}$ &    0.0045 &  0.0984 &  1.0185\\
${\rm MAE}_{\alpha_{\rm LW}}$ &    0.0524&  0.1651 &   0.5096\\
${\rm MAE}_{M_{\rm cool0}}$ &   \---&  0.1665 &   0.6054\\
$R^2_{\alpha_{\rm LW}}$ &  0.9961 & 0.9497 & 0.5355 \\
$R^2_{M_{\rm cool0}}$ &   \--- & 0.9470 & 0.4081 \\
\hline
\end{tabular}
\label{tab:cases}
\end{table*}

\section{Discussion}
\label{sec:discussion}

The results presented above demonstrate that VAO features provide a physically motivated probe of the cooling threshold for Pop III star formation and, in turn, the efficiency of LW feedback. In this section we discuss the main modeling assumptions, observational limitations, and possible extensions of our work.

\subsection{Parameterization of the Cooling Threshold}

We adopt a two-parameter formula of the critical halo mass for H$_2$ cooling in the presence of LW radiation \citep{machacekSimulationsPregalacticStructure2001} and dark matter-baryon relative streaming motion \citep{fialkovImpactRelativeMotion2012}. 
In particular, we assume: (i)
		A redshift-independent baseline cooling threshold $M_{\rm cool0}$; (ii)
A power-law LW dependence controlled by $\alpha_{\rm LW}$; (iii)
	A fixed streaming suppression coefficient $\alpha_{v_{\rm db}}$.
This parameterization captures the leading physical effects while keeping the inference problem tractable. However, it is obviously not unique. In Appendix \ref{sec:vary_alpha_v_db} we show how the VAO wiggles change for various $\alpha_{v_{\rm db}}$.

In more detailed treatments, the baseline cooling threshold may evolve explicitly with redshift (e.g., in \citet{visbalHighredshiftStarFormation2014} they use $M_{\rm cool0}\propto(1+z)^{-1.5}$) or depend on additional physical parameters such as the local gas chemistry, halo assembly history. 
For $M_{\rm cool2}$, there are more complicated parameterizations derived from simulations involving both the LW feedback and the relative streaming motion, for example \citet{kulkarniCriticalDarkMatter2021,schauerInfluenceStreamingVelocities2021}. Incorporating such refinements would increase the dimensionality of parameter space and likely enhance degeneracies. 

Importantly, our results suggest that VAO features primarily constrain the evolution of the effective cooling mass during the VAO-active redshift interval, rather than the instantaneous value of any single parameter. This explains the degeneracy direction identified in Sec. \ref{sec:results}: combinations of $M_{\rm cool0}$ and $\alpha_{\rm LW}$ that produce similar cooling-mass evolution during $z \sim 17-21$ yield nearly indistinguishable VAO signals.
While a more complex parameterization may shift quantitative constraints, our qualitative conclusion is robust:
VAO amplitudes are sensitive to whether the cooling threshold lies within the streaming-suppressed halo regime.

\subsection{The net wiggles for using polynomials with different degrees} \label{sec:degree_poly}

Since the dark matter-baryon relative streaming motion not only generates the wiggles, but also changes the shape and amplitude of the smooth component, the stability of the method for extracting the net wiggles is quite important. If the method is not stable, it may generate artificial frequency-dependence (evolution trend) on the net VAO wiggles, and confuse the parameter inference. 

\begin{figure}
    \centering
        \includegraphics[width=0.8\linewidth]{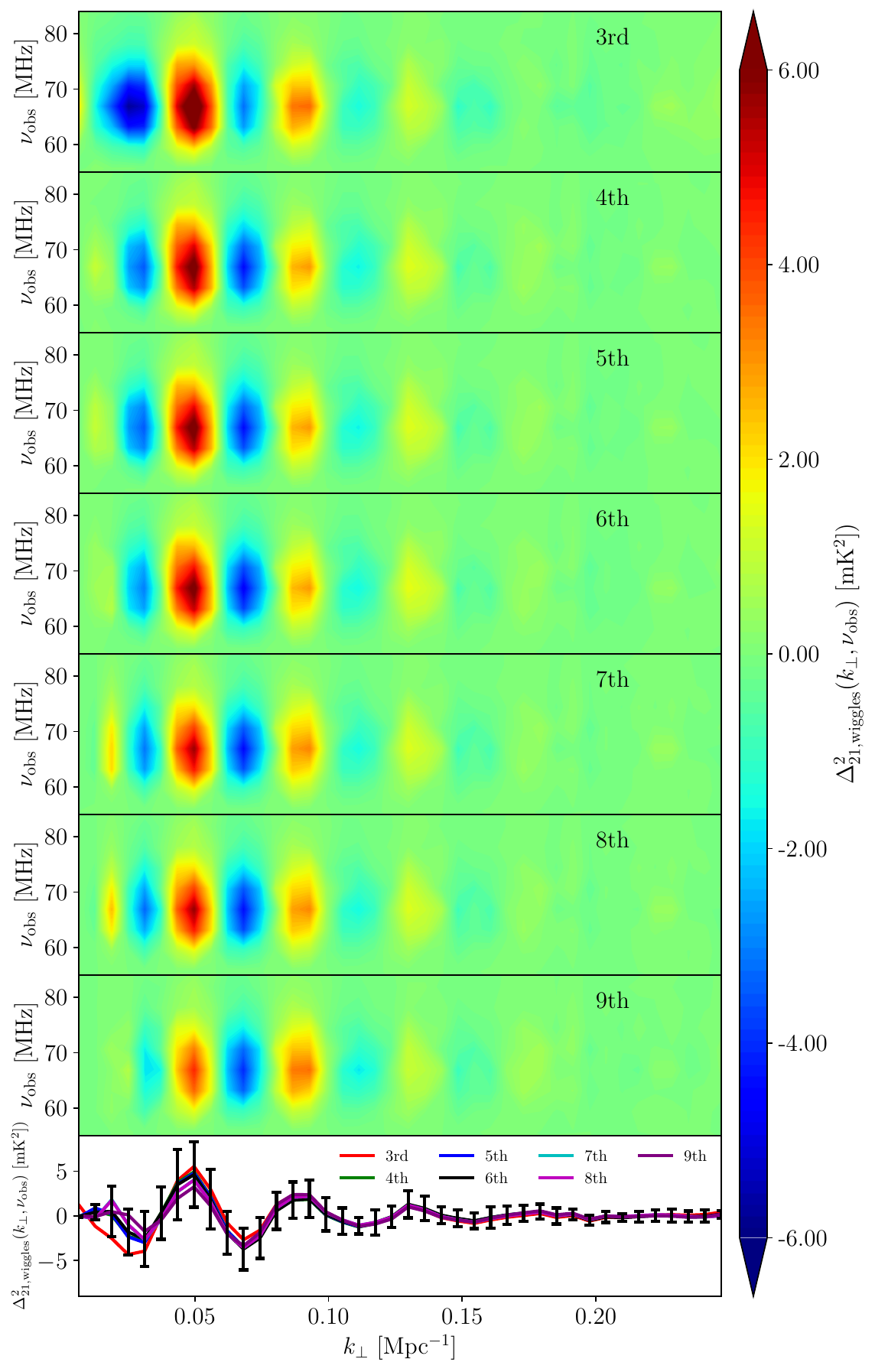 }
    \caption{{\it Top 7 panels:}
    The net VAO wiggles obtained by using polynomials with different degrees to fit the smooth component in the MAPS, for the model with $M_{\rm cool0}=10^5~M_\odot$ and $\alpha_{\rm LW}=0.0$. {\it Bottom panel:} the wiggles at $\nu_{\rm obs}=70$ MHz. Errorbars are standard deviations of the wiggles among the 36 realizations.
    }
    \label{fig:poly_degrees}
\end{figure}

In Fig. \ref{fig:poly_degrees} we show the net VAO wiggles for using polynomials with degrees ranging from 3 to 9 to fit the smooth component in the MAPS with $M_{\rm cool0}=10^5~M_\odot$ and $\alpha_{\rm LW}=0.0$. We see that, for 4th to 7th degree polynomials, the results are almost identical. For 3rd degree polynomial, the wiggles are slightly higher, while for 8th and 9th degree polynomials, the wiggles are slightly lower. Nevertheless, the discrepancies are much smaller than the cosmic variance. We therefore conclude that, at least for the mock MAPS samples in our paper, using 6th degree polynomial to fit the smooth component is stable enough.

\subsection{Foregrounds and Large-Scale Modes}

Since the VAO wiggles are large-scale features, precisely where foreground contamination is most severe, this presents a major observational challenge.

 Generally, in the $k_\parallel-k_\perp$ plane, the modes in the wedge with 
\begin{equation}
k_\parallel  \lesssim k_{\parallel,\rm wedge} =k_\perp \sin(\theta_{\rm FoV})\left( \frac{D_{\rm M}(z)E(z)}{c/H_0(1+z)} \right)
\end{equation}
would be contaminated (\citealt{dattaBRIGHTSOURCESUBTRACTION2010,moralesFOURFUNDAMENTALFOREGROUND2012,liuEpochReionizationWindow2014,jensenWedgeBiasReionization2016}), 
where $D_{\rm M}$ is the transverse comoving distance, $E(z)=\sqrt{\Omega_m(1+z)^3+\Omega_\Lambda}$, $c$ is the light speed and $H_0$ is the Hubble constant. Since VAO wiggles originate from large-scale correlations, aggressive wedge removal can significantly reduce signal amplitude.

\begin{figure}
    \centering
        \includegraphics[width=\linewidth]{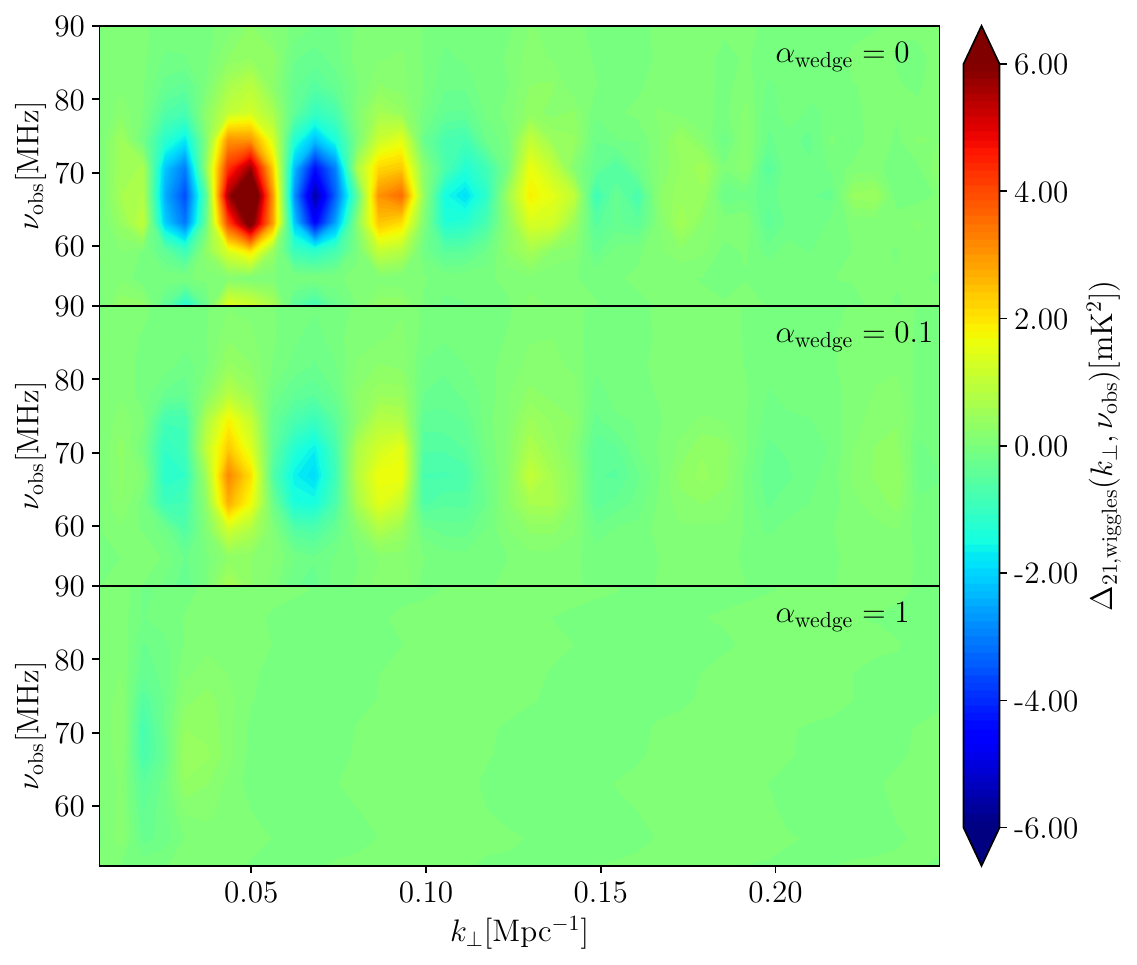 }
    \caption{
    The VAO features if we remove all modes with $k_\parallel < \alpha_{\rm wedge}k_{\parallel,\rm wedge}$. For comparsing purpose, in the top panel we show the results if all modes are kept ($\alpha_{\rm wedge}=0$). 
    }
    \label{fig:FG_wedge}
\end{figure}

Our analysis assumes optimistic foreground mitigation, consistent with standard sensitivity forecasts (e.g.\ the  optimistic setup in SKA sensitivity calculators like the {\tt 21cmSense} \citep{Pober2013AJ,Pober2014ApJ,Murry2024JOSS}). When progressively removing low-$k_\parallel$ modes, we find that: (i) Complete wedge removal strongly suppresses VAO visibility (see the bottom panel of Fig. \ref{fig:FG_wedge}); (ii) Partial removal (e.g., $k_\parallel < 0.1\,k_{\parallel,\rm wedge}$) reduces but does not erase the signal (see the middle panel of Fig. \ref{fig:FG_wedge}).

This indicates that meaningful constraints on LW feedback require foreground cleaning beyond conservative wedge avoidance strategies. However, VAO features possess an important advantage: their oscillatory structure appears at cosmologically fixed transverse scales. Foregrounds, which are spectrally smooth and lack coherent acoustic structure, are unlikely to mimic these wiggles at the correct scales. This distinctive morphology may aid in separating signal from residual foreground contamination.
A full assessment of foreground systematics using dedicated end-to-end simulations is beyond the scope of this work.

\subsection{Noise and Observational Requirements}

Including instrumental noise corresponding to the SKA-low AA* configuration substantially degrades parameter recovery. For integration times of $10^4$ hours, the LW efficiency can be constrained only at the level of broad parameter ranges, while the baseline cooling threshold becomes difficult to recover independently.

This requirement should be interpreted cautiously, as we have assumed Gaussian thermal noise, neglected calibration systematics and foreground residuals, and assumed a single-field observation.
Under these optimistic assumptions, $10^4$ hours already represents a demanding observational program. More realistic systematics would likely increase the required integration time or necessitate multi-field averaging.

Nevertheless, the scaling behavior we found is encouraging: constraints improve rapidly in the regime where the VAO signal amplitude is large (small $\alpha_{\rm LW}$ and low $M_{\rm cool0}$). Thus, even partial detections could potentially distinguish between strong and weak LW feedback scenarios.

\subsection{Choice of Estimator: MAPS vs.\ Three-dimensional Power Spectrum}

\begin{figure}
    \centering
        \includegraphics[width=0.9\linewidth]{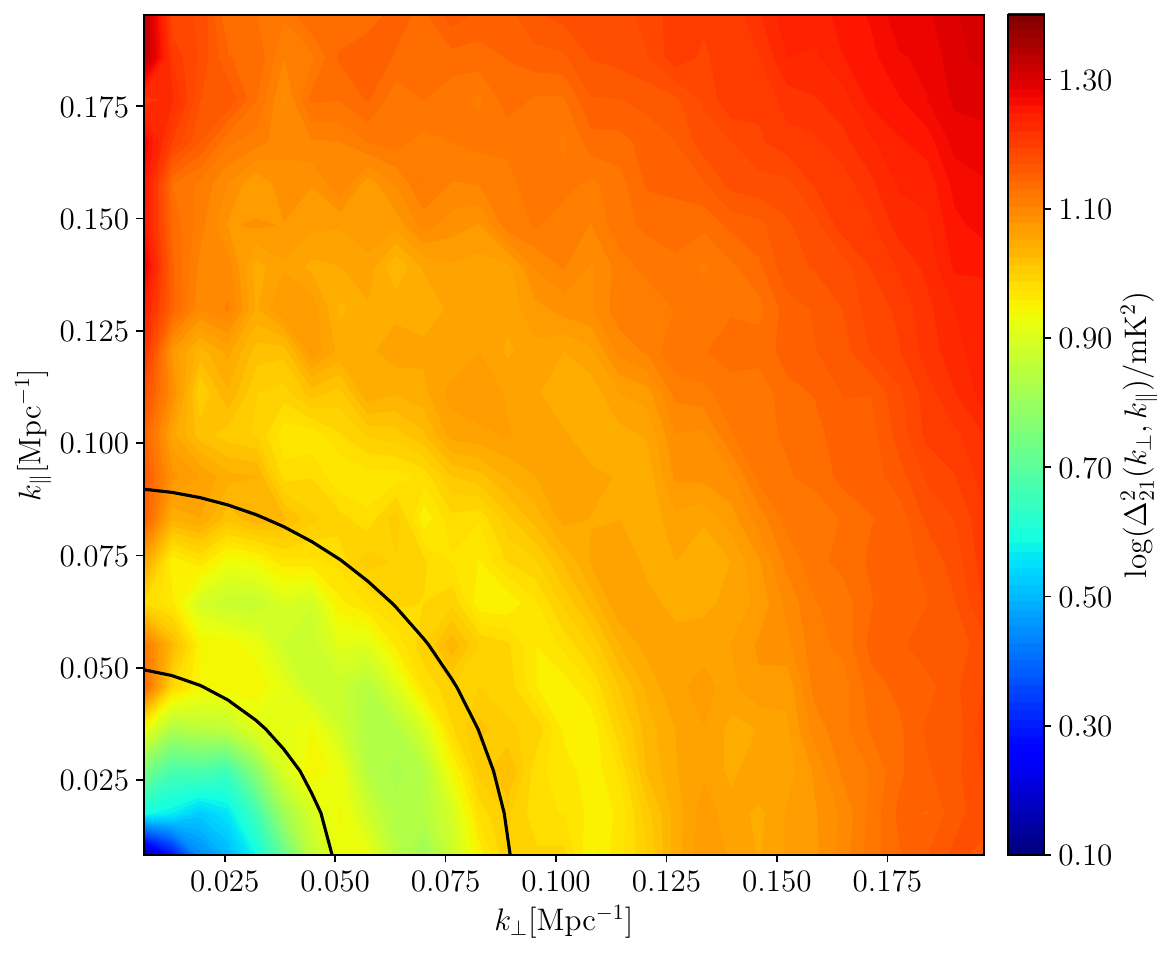}
        \includegraphics[width=0.9\linewidth]{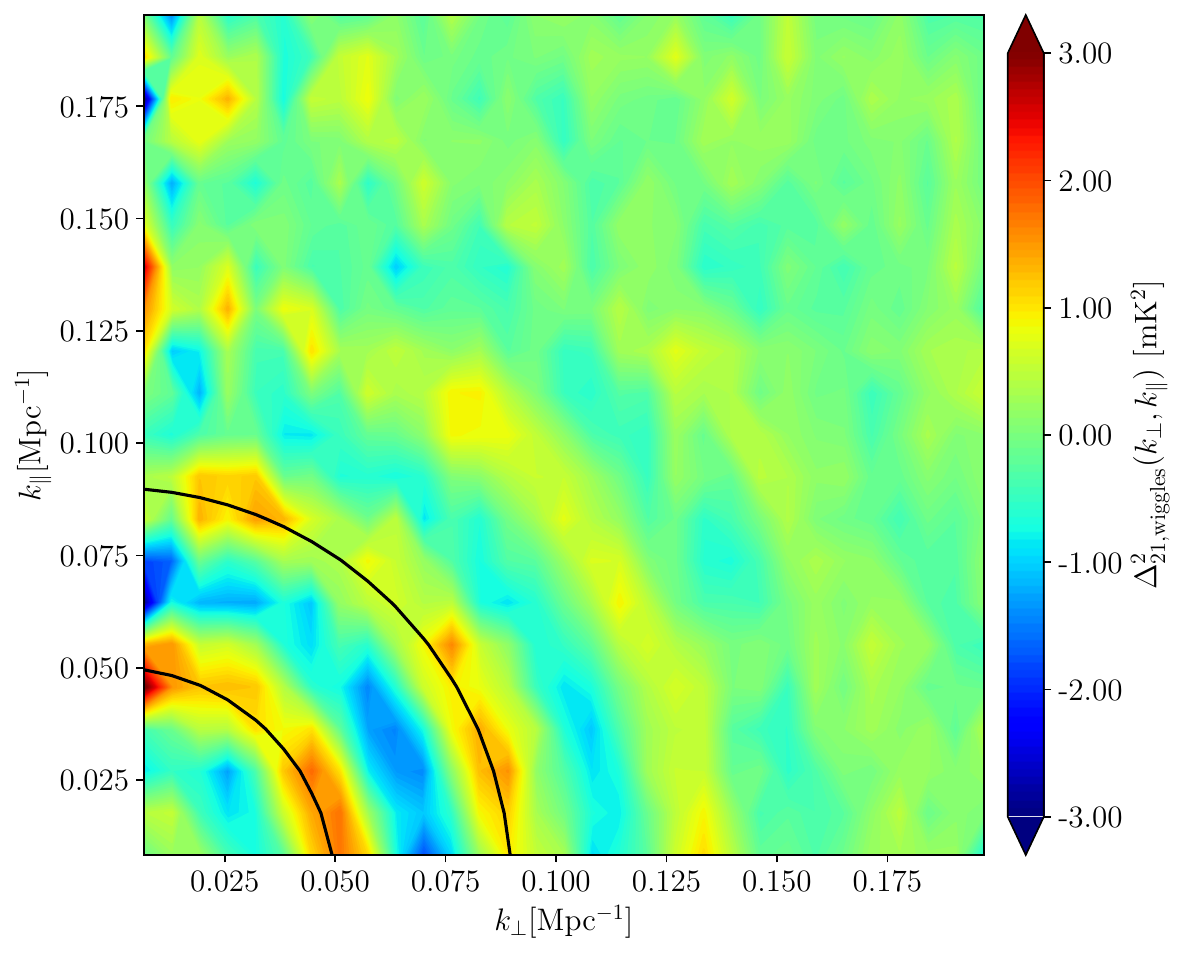}
    \caption{
    {\it Top:} The cylinder power spectrum $\Delta^2_{21}(k_\perp,k_\parallel)$ of the lightcones from the simulation with $M_{\rm cool0}=10^5~M_\odot$ and $\alpha_{\rm LW}=0$. Here is the mean result of 36 realizations.
    {\it Bottom:} the VAO wiggles $\Delta^2_{21,\rm wiggles}(k_\perp,k_\parallel)$ in the above panel. To guide eyes we mark the locations of the first and second VAO peaks at  $\sqrt{k_\perp^2+k_\parallel^2}\approx 0.05$ and 0.09 Mpc$^{-1}$ by solid curves. 
    }
    \label{fig:cylinder_PS}
\end{figure}

Because of the lightcone effect, $P_{21}(k)$ is not suitable for measuring the VAO features, so we used the MAPS in this paper. In principle the cylinder power spectrum $P_{21}(k_\perp,k_\parallel)$ also contains the lightcone effect, and it provides more information of the correlation along line of sight than MAPS. In Fig. \ref{fig:cylinder_PS} we show the $\Delta^2_{21}(k_\perp,k_\parallel)$ and the residual wiggles by subtracting a two-dimensional polynomial fitting, for the simulation with $M_{\rm cool0}=10^5~M_\odot$ and $\alpha_{\rm LW}=0$. We found that the wiggles in the $k_\parallel$ direction are not as obvious as in the $k_\perp$ driection, not only because the lightcone effect (the VAO features only exist in a narrow length scale), but also because in the $k_\parallel$ direction the number of independent Fourier modes is much smaller than in the $k_\perp$ direction, so the sample variance must be larger. For this reason, we believe that using  $P_{21}(k_\perp,k_\parallel)$ will not improve the results of our paper. However, it is interesting to note that, for the same simulation, the amplitude of the VAO wiggles is stronger for the MAPS than for the cylinder power spectrum, see the first column of Fig. \ref{fig:meanVAO} and Fig. \ref{fig:cylinder_PS}.
An explicit dedicated exploration for this will be warranted in the future.

\subsection{Role of Machine Learning}

The mapping between cooling-threshold parameters and VAO morphology is highly nonlinear and partially degenerate. While simpler summary statistics (e.g., peak amplitude evolution) may capture part of the information, the CNN efficiently exploits the full two-dimensional structure of the MAPS, including correlations across frequencies and wavenumbers.

Importantly, the CNN does not create new information; it compresses the available morphological structure into parameter estimates. The degeneracy patterns recovered by the CNN reflect genuine physical degeneracies in the cooling-mass evolution, rather than artifacts of the inference method.

In Appendix \ref{ap:fisher} we compare the predicted $\alpha_{\rm LW}$-$M_{\rm cool0}$ constraints by Fisher matrix with the CNN. The Fisher matrix has looser constraints then CNN, so even in this two-parameter model, the CNN has much better performance. 

Future work could incorporate Bayesian neural networks to obtain full posterior distributions, or compare CNN-based inference with traditional likelihood-based approaches built on physically motivated summary statistics.

\section{Summary  } 
\label{sec:summary}

In this paper, we have investigated whether Velocity Acoustic Oscillation (VAO) features in the Cosmic Dawn 21 cm signal can be used to constrain Lyman–Werner (LW) feedback on Pop III stars formation. The central idea is simple: VAO features are strongest when star formation occurs in minihalos that are sensitive to dark matter–baryon relative streaming velocities. If LW feedback raises the cooling threshold above this regime, the VAO signal is suppressed. The presence, absence, and evolution of VAO wiggles therefore encode information about the cooling physics of the first stars.

Using semi-numerical lightcone simulations that include H$_2$ cooling, LW feedback, and streaming suppression, we analyzed the multi-frequency angular power spectrum (MAPS) and trained a convolutional neural network (CNN) to infer two parameters: the baseline cooling threshold $M_{\rm cool0}$ and the LW feedback efficiency $\alpha_{\rm LW}$.

Our main findings are:
	(i)	\textit{VAO wiggles are a sensitive probe of the effective cooling mass.}
In noise-free simulations, both $M_{\rm cool0}$ and $\alpha_{\rm LW}$ can be recovered with high accuracy. The constraints arise from the redshift evolution of the VAO amplitudes rather than from any single snapshot.
	(ii)	\textit{Degeneracies reflect similar cooling-mass histories.}
The dominant parameter degeneracy corresponds to combinations of $M_{\rm cool0}$ and $\alpha_{\rm LW}$ that produce nearly identical evolution of the effective cooling threshold during the VAO-active redshift interval. This indicates that VAO measurements primarily constrain the time-dependent cooling mass, rather than individual model parameters in isolation.
	(iii)	\textit{Lightcone evolution matters.}
Since the 21 cm signal evolves rapidly during Cosmic Dawn, preserving frequency information is essential. The MAPS estimator retains VAO structure more effectively than spherically averaged three-dimensional power spectra.
	(iv)	\textit{Observational requirements are demanding.}
When SKA-low AA* noise is included, meaningful constraints require integration times of order $10^4$ hours under optimistic assumptions. Foreground wedge removal further reduces sensitivity to the large-scale modes that host VAO features. While challenging, these requirements are not fundamentally prohibitive if foreground mitigation and calibration improve beyond conservative limits.

The key conceptual result of this work is that VAO wiggles provide a physically clean diagnostic of whether star formation during Cosmic Dawn occurred in streaming-sensitive minihalos. Because the acoustic scale is set by well-understood early-Universe physics, the interpretation of the signal is comparatively robust against astrophysical modeling uncertainties.

A future detection of VAO features would therefore simultaneously confirm the role of streaming velocities in early structure formation and provide quantitative constraints on LW-regulated H$_2$ cooling. Conversely, the absence of VAO wiggles at the expected amplitude would imply that LW feedback (or other feedback processes) efficiently suppressed star formation in low-mass halos.

In this sense, VAO measurements offer a direct observational pathway to probing the cooling threshold of the first stars — a quantity that is otherwise inaccessible.

\normalem
\begin{acknowledgements}
We thank the anonymous referee very much for the constructive suggestions that help to improve our paper. This work is supported by the NSFC International (Regional) Cooperation and Exchange Project No. 12361141814, the National SKA Program of China Nos. 2020SKA0110402 \& 2020SKA0110401, the China's Space Origins Exploration Program Nos. GJ11010405 \& GJ11010401, and the Project Supported by the Specialized Research Fund for State Key Laboratory of Radio Astronomy and Technology. It uses the computing resources of the National Supercomputing Center in Tianjin. EDK acknowledges
 support from the U.S.-Israel Bi-national Science
Foundation (NSF-BSF grant 2022743 and BSF grant 2024193) and the Israel National Science Foundation (ISF grant 3135/25), as
well as support from the joint Israel-China   program (ISF-NSFC grant  3156/23). 

\end{acknowledgements}

\appendix

\section{The resolution and boxsize}\label{sec:res_boxsize}

\begin{figure*}
    \centering
\includegraphics[width=\linewidth]{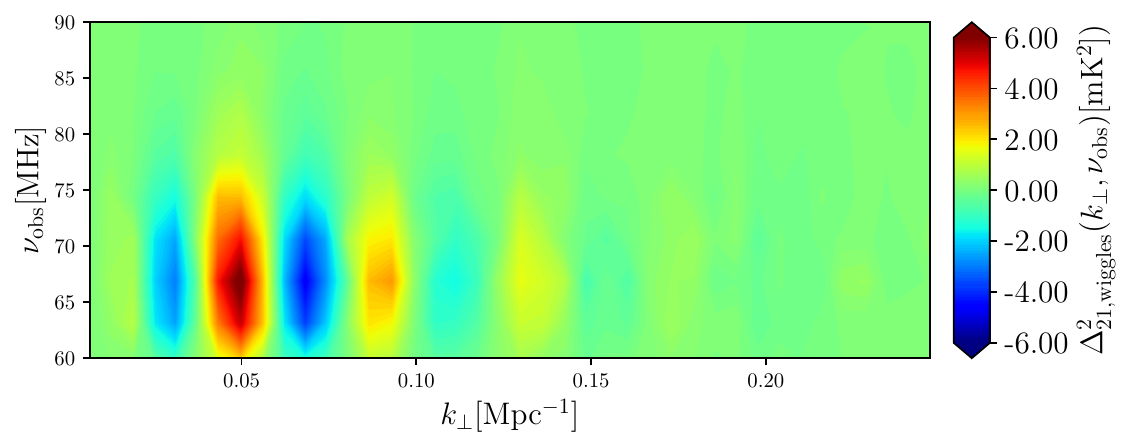}
\includegraphics[width=\linewidth]{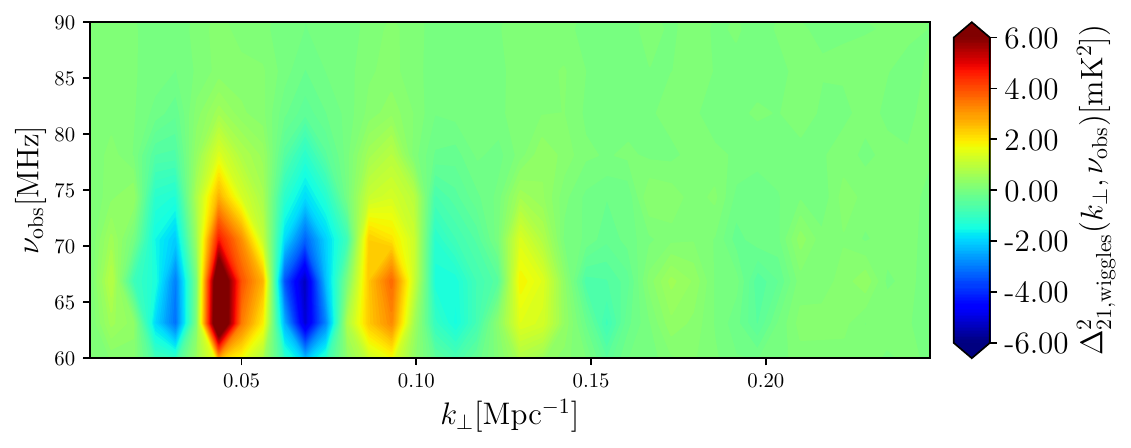}
\includegraphics[width=\linewidth]{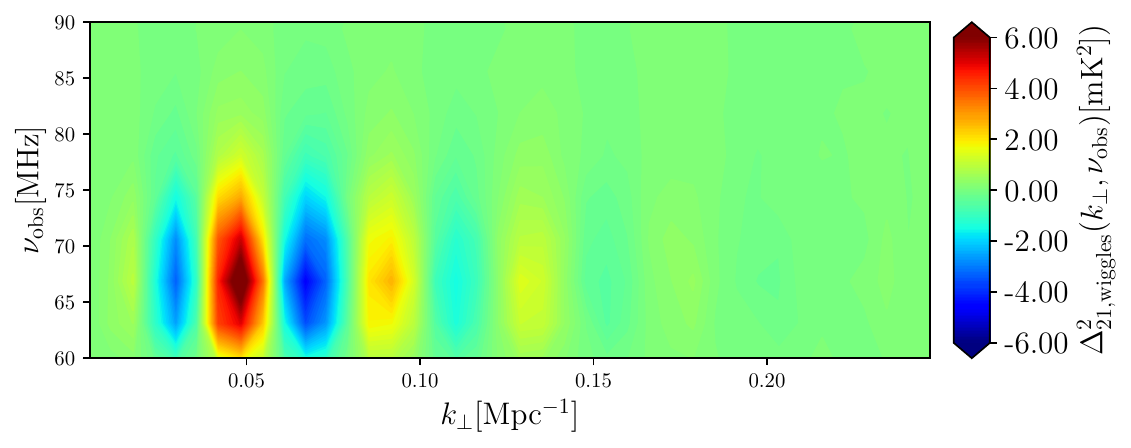}
    \caption{The MAPS maps for different box size and resolution. {\it Top:} our fiducial box size $1800$ Mpc and cells $300^3$. {\it Middle:} the box size is the same to fiducial one but the cells are $600^3$. {\it Bottom:} the cells are similar to middle panel but the box size is 3600 Mpc. For all panels $M_{\rm cool0}=10^5~M_\odot$ and $\alpha_{\rm LW}=0$.
    }
    \label{fig:resolution_box}
\end{figure*}

In Fig. \ref{fig:resolution_box} we show the MAPS maps for higher resolution  or larger box size, compared with the fiducial one. We see that basically the results are consistent with each other. However if the resolution is higher and/or the box size is larger, the signal is a bit larger. This will strengthen our conclusion.

\section{The effect of varying $\alpha_{v_{\rm db}}$ }\label{sec:vary_alpha_v_db}

\begin{figure*}
\centering
\includegraphics[width=0.45\linewidth]{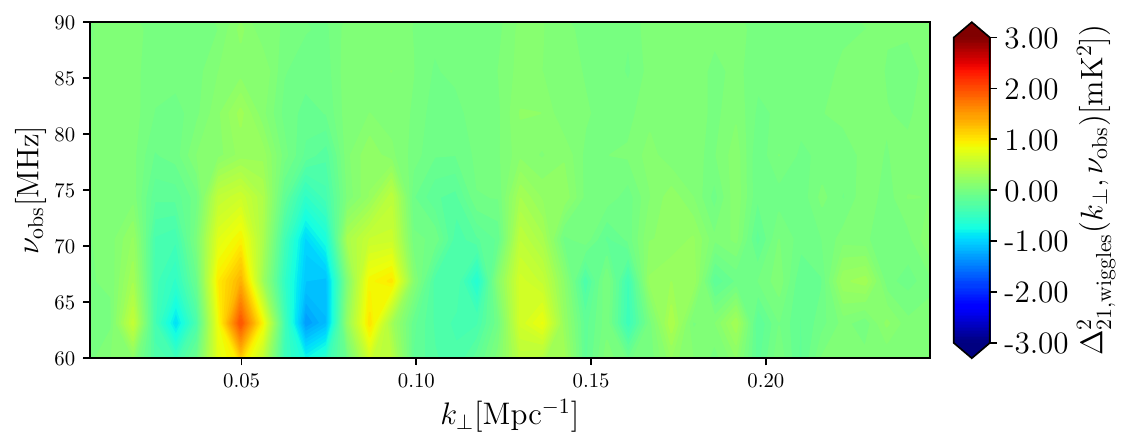}
\includegraphics[width=0.45\linewidth]{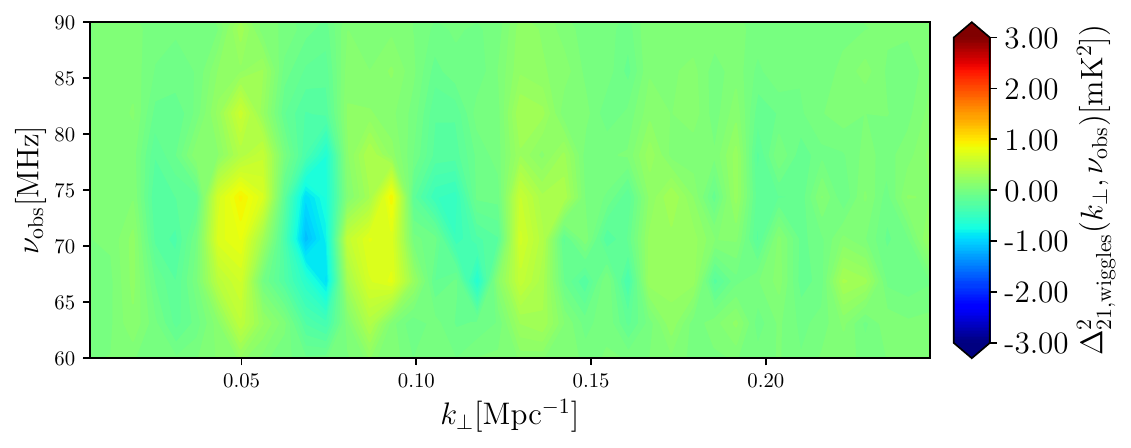}
\includegraphics[width=0.45\linewidth]{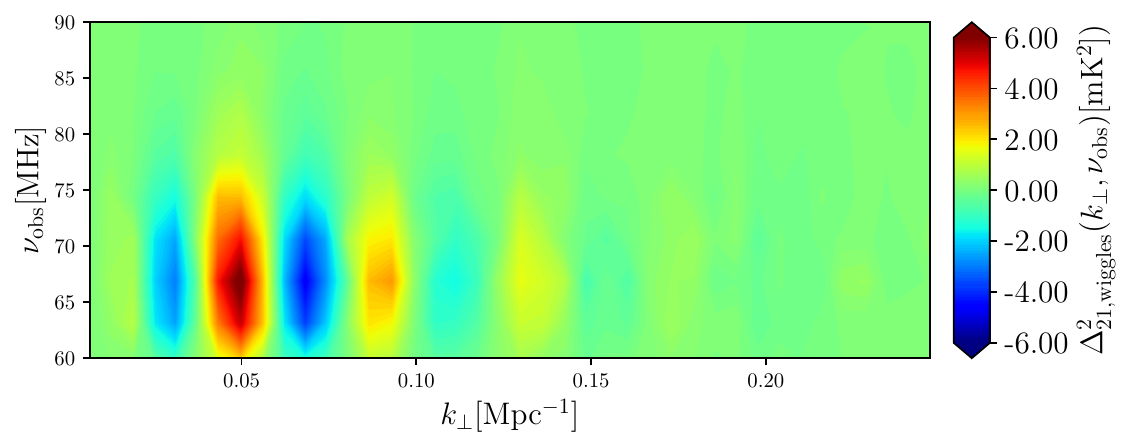}
\includegraphics[width=0.45\linewidth]{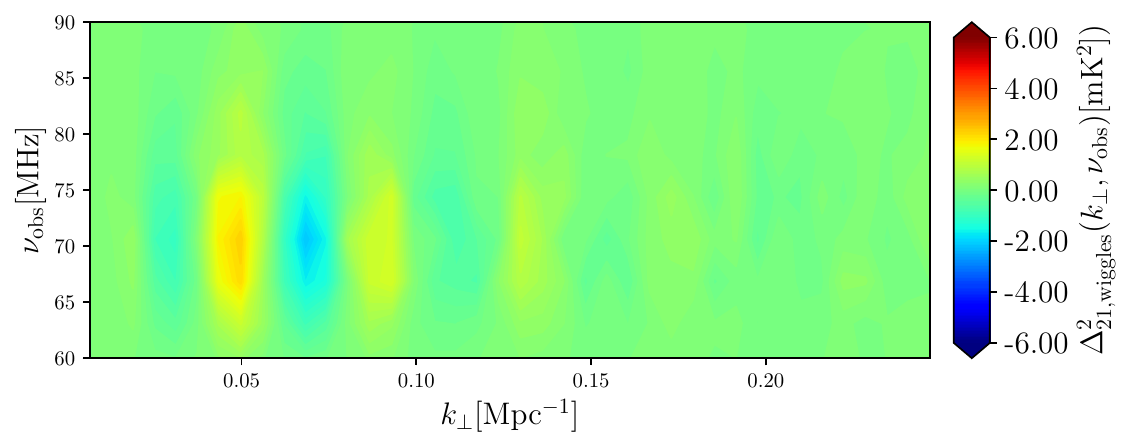}
\includegraphics[width=0.45\linewidth]{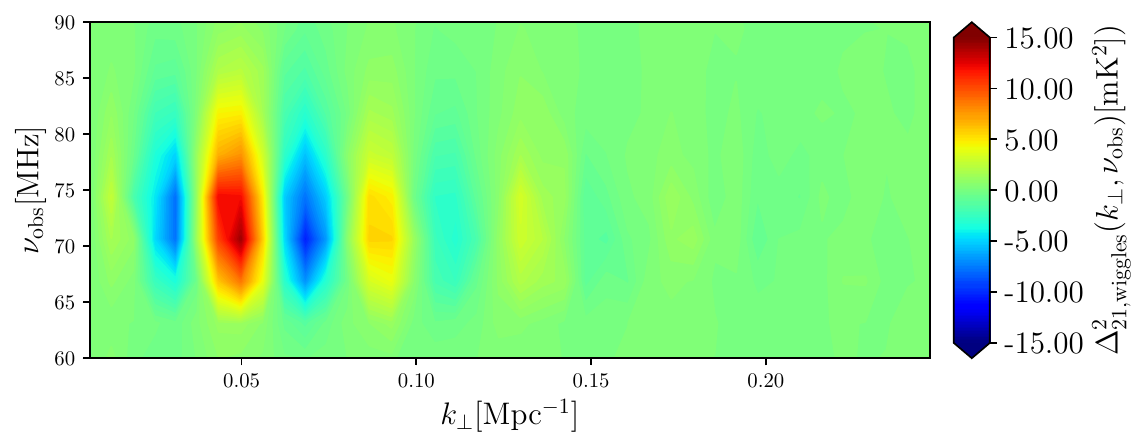}
\includegraphics[width=0.45\linewidth]{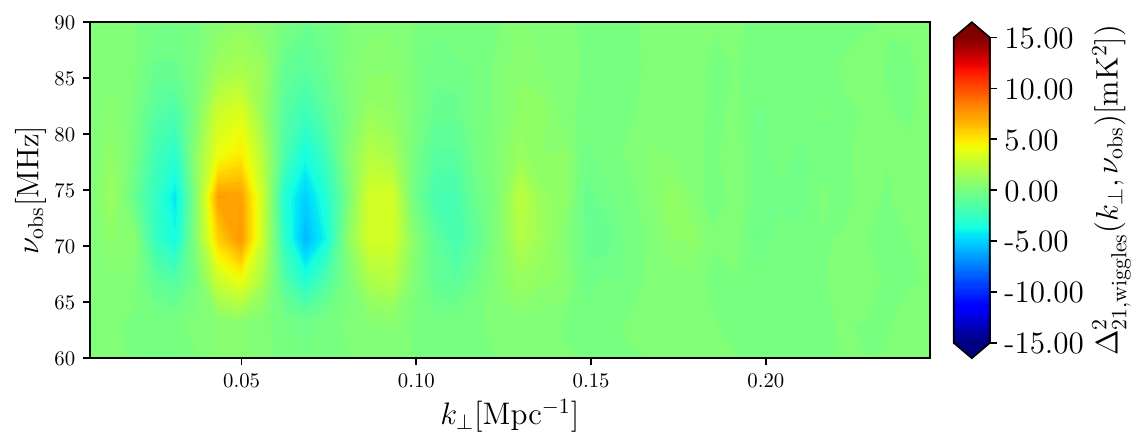}
    \caption{The VAO wiggles for different $\alpha_{\rm LW}$ and $\alpha_{v_{\rm db}}$. For display purpose we only show the frequency range $60~{\rm MHz} \le \nu_{\rm obs} \le 90~{\rm MHz}$ where there are VAO wiggles visible. {\it Left column:} $\alpha_{\rm LW}=0$. {\it Right column:} $\alpha_{\rm LW}=2.0$. In each column, from top to bottom, the panel corresponds to $\alpha_{v_{\rm db}}=2.0$, 4.015 (the fiducial model) and 8.0 respectively.
    }
    \label{fig:VAO_alpha_v_db}
\end{figure*}

In Fig. \ref{fig:VAO_alpha_v_db} we show the VAO wiggles for various $\alpha_{v_{\rm db}}$. We see that the larger the $\alpha_{v_{\rm db}}$, the stronger the VAO wiggles. Moreover, larger $\alpha_{v_{\rm db}}$ delays the signal, so VAO wiggles are shifted to higher frequency. While, although in principle stronger LW feedback also delays the VAO wiggles and shift them to higher frequency, it suppresses the VAO wiggles at the same time. For this reason, in principle $\alpha_{v_{\rm db}}$ does not necessarily  degenerate with $\alpha_{\rm LW}$, and there could be chance to infer them both if $\alpha_{v_{\rm db}}$  is also involved.

\section{The predicted constraints on $\alpha_{\rm LW}$ and $M_{\rm cool0}$ by Fisher matrix }\label{ap:fisher}

\begin{figure}
    \centering
        \includegraphics[width=0.9\linewidth]{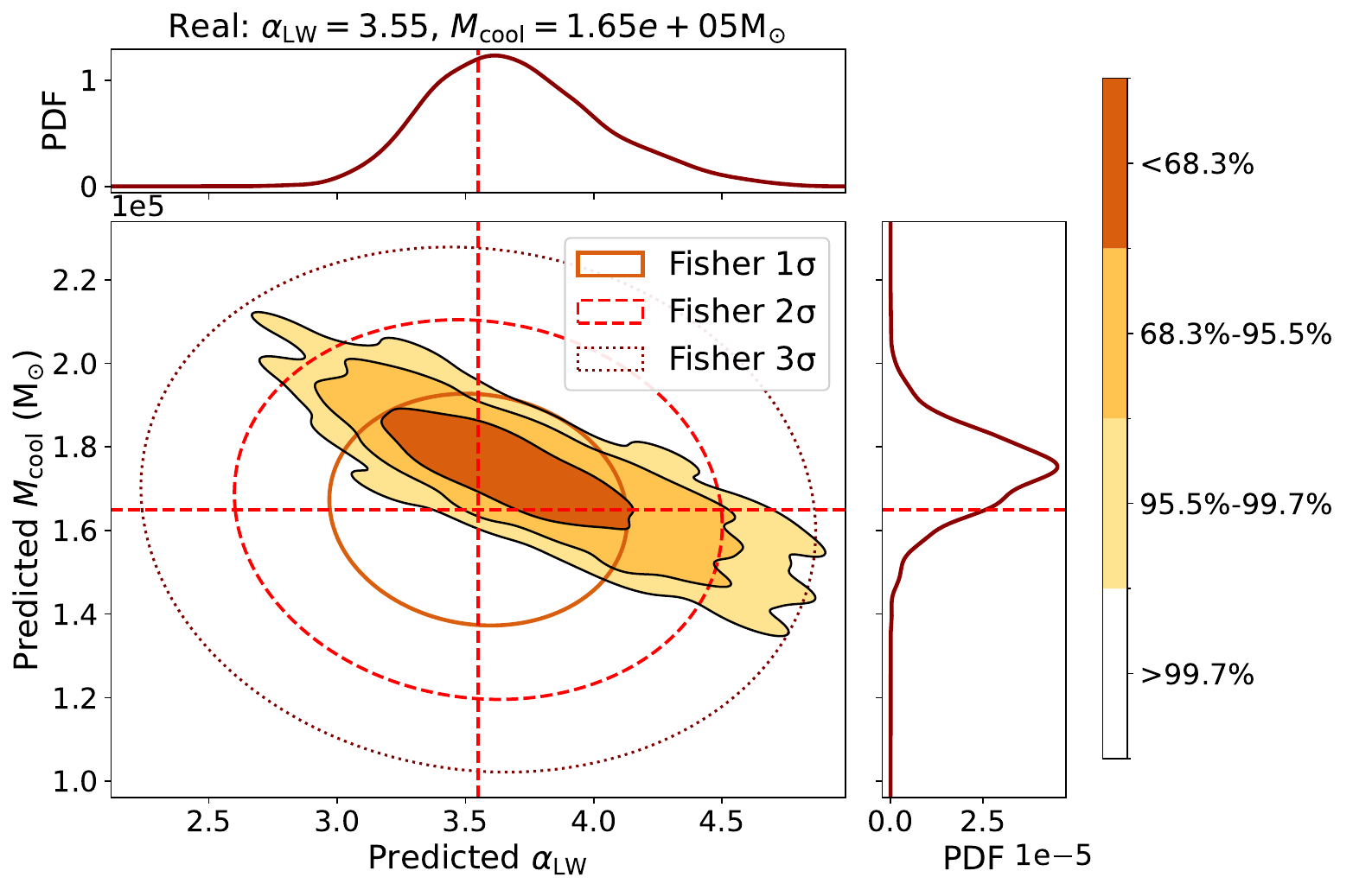}
    \caption{We add the contour predicted by Fisher matrix to Fig. \ref{fig:joint_PD}.
    }
    \label{fig:joint_PD_fisher}
\end{figure}

In Fig. \ref{fig:joint_PD_fisher} we add the forecast of the $\alpha_{\rm LW}$-$M_{\rm cool0}$ constraints, given by Fisher matrix \citep{Coe_2009arXiv0906.4123C}. We calculate the Fisher matrix by numerical derivation. To avoid the contamination we only use the MAPS map in the range with $65~{\rm MHz}\lesssim \nu_{\rm obs}\lesssim 90~{\rm MHz}$, because the pixel values of the MAPS map outside this range are very close to zeros and may introduce numerical errors. 

From Fig. \ref{fig:joint_PD_fisher}, we see that basically the Fisher matrix gives comparable constraint on $\alpha_{\rm LW}$ as CNN. However, the constraint on $M_{\rm cool0}$ is much looser than CNN, therefore the parameter degeneracy direction is not as obvious as the CNN. This figure highlights the value of the CNN even in the two-parameter model.

\bibliographystyle{raa}
\bibliography{bibtex}

\end{document}